\documentclass[aps, pra, twocolumn, superscriptaddress, amsmath,  tightenlines, longbibliography]{revtex4-1}

\usepackage{tikz}

\usepackage{graphicx} 

\usepackage{epstopdf}
\usepackage{dcolumn}
\usepackage{graphicx}
\usepackage{mathrsfs}
\usepackage{subfigure}
\usepackage{booktabs}
\usepackage{amsmath}
\usepackage{physics}
\usepackage{dsfont}
\usepackage{amstext}
\usepackage{amssymb}
\usepackage{amsbsy}
\usepackage{bbm}
\usepackage{amsthm}
\usepackage{color}
\usepackage[colorlinks,citecolor=blue]{hyperref}

\usepackage{url}
\usepackage[colorlinks]{hyperref}
\hypersetup{%
	plainpages=true,
	breaklinks=true,       
	hypertexnames=false,  
	pageanchor=true,
	colorlinks=true,
	linkcolor={blue},
	citecolor={red},
	urlcolor={blue},
	anchorcolor={black}
}

 \makeatletter

\newcommand{\Rmnum}[1]{\expandafter\@slowromancap\romannumeral #1@}
\makeatother

\begin{document}


\title{Quantum Simulation of Bound-State-Enhanced Quantum Metrology}
	\author{Cheng-Ge Liu}
    \thanks{These authors contributed equally to this work.}
    \affiliation{%
		Department of Physics, Applied Optics Beijing Area Major Laboratory, Beijing Normal University, Beijing 100875, China
	}%
    \author{Cong-Wei Lu}
    \thanks{These authors contributed equally to this work.}
    \affiliation{%
		Department of Physics, Applied Optics Beijing Area Major Laboratory, Beijing Normal University, Beijing 100875, China
	}%
    \author{Na-Na Zhang}
    \affiliation{%
		School of Optoelectronics Engineering, Chongqing University of Posts and Telecommunications, Chongqing 400065, China
	}%
	\author{Qing Ai}
	\email{aiqing@bnu.edu.cn}
	\affiliation{%
		Department of Physics, Applied Optics Beijing Area Major Laboratory, Beijing Normal University, Beijing 100875, China
	}%




\date{\today}

\begin{abstract}
Quantum metrology explores quantum effects to improve the measurement accuracy of some physical quantities beyond the classical limit. However, due to the interaction between the system and the environment, the decoherence can significantly reduce the accuracy of the measurement. Many methods have been proposed to restore the accuracy of the measurement in the long-time limit. Recently, it has been found that the bound state can help improve measurement accuracy and recover the $t^{-1}$ scaling [K. Bai, Z. Peng, H. G. Luo, and J. H. An,
Phys. Rev. Lett. 123, 040402 (2019)]. Here, by using $N$-qubits, we propose a method to simulate the open quantum dynamics of the hybrid system including one atom and coupled resonators. We find that the error of the measurement can reduce as the time increases due to the existence of the bound state. By both analytical and numerical simulations, we prove the $t^{-1}$ scaling of the measurement error can be recovered when there is a bound state in the hybrid system. Interestingly, we observe that there are regular
oscillations  which can be used for the evaluation of the atomic transition frequency. For a finite-$N$, the duration of the regular
oscillations doubles as one more qubit is involved.
\end{abstract}

\maketitle 

\section{Introduction} 
Compared with classical metrology, quantum metrology can greatly improve the measurement accuracy and play a significant role in  gravitational-wave detection \cite{grote2013PRL,schnabel2010NC,ligo2011NP} and quantum radar \cite{barzanjeh2015PRL,maccone2020PRL,arrad2014PRL}, atomic clocks \cite{kruse2016PRL,hosten2016Nat,pezze2020PRL,kaubruegger2021PRX}, magnetometers \cite{thiel2016NN,taylor2008NP,bao2020Nature}, gravimeters \cite{hardman2016PRL,asenbaum2017PRL}, navigation and biological monitoring \cite{cai2013PRL,taylor2016PR,crespi2012APL,taylor2014PRX,cai2012PRA,yang2012PRA}, quantum biology \cite{taylor2014PRX,crespi2012APL,taylor2016PR} and so on.

According to the central-limit theorem \cite{Giovannetti2011NP,Wang2017NJP}, by performing a large number of measurements, the error of the measurements will be reduced by a factor $\sqrt{M}$ with $M$ being the number of measurements, i.e., the shot-noise limit (SNL) or the standard quantum limit \cite{Wang2017NJP}. The quantum metrology explores quantum entanglement \cite{szigeti2020PRL,luo2017PRA,pezze2009PRL}, coherence \cite{joo2011PRL,he2021PRA,zhou2018NC} and squeezing \cite{ma2011PR,engelsen2017PRL,holland1993PRL} to improve the accuracy of the measurement in order to reach the Heisenberg limit (HL) which scales as $M^{-1}$. As long as the system-bath interaction is present, the measurement of a specific physical quantity is inevitably affected by the error. The precision of the quantum metrology in a Markovian bath is reduced the SNL \cite{Huelga1997PRL}. The SNL can also be defeated in non-Markovian noise and thus achieve the Zeno limit (ZL), i.e., $M^{-3/4}$ \cite{chin2012PRL,Matsuzaki2011PRA,long2022PRL}.


Ramsay spectrum is widely used in practical quantum metrology, which approaches the HL when the noise is absent \cite{bai2023PRL}. However, in practice, since the system is open to the environment, the measurement accuracy of physical quantities will be affected by environmental noise. In the presence of Markovian pure-dephasing noise,the accuracy is reduced from the HL back to the SNL \cite{Huelga1997PRL}. Although under non-Markovian noises the accuracy can reach the ZL, it is still lower than the HL \cite{Matsuzaki2011PRA,long2022PRL,smirne2016PRL,Macieszczak2015PRA}. In both cases, the measurement error will diverge over time. In order to improve the measurement accuracy, many methods have been proposed, such as purification \cite{Yamamoto2022PRL}, error correction \cite{Dur2014PRL,Rojkov2022PRL,reiter2017NC}, non-degenerate measurements \cite{Rossi2020PRL}, and bound states outside continuum \cite{bai2019PRL,Wang2017NJP}. None of these proposals recover the HL and especially fail to resolve the issue of error diverging over time. It is natural to ask the question how to restore the measurement error without diverging over time?

Recent studies have shown that when there is a bound state in the open quantum system, it allows the measurement precision to beat the SNL and recover the ZL over long encoding times \cite{bai2019PRL}. The bound state is an interesting state that can be observed in the boson-impurity model in which two-level systems are coupled to a boson bath. Here, the impurity is the emitter (e.g. atom) and the boson bath is the electromagnetic field mode \cite{Shi2016PRX}. Bound states can possess many interesting phenomena, such as fractional decay, localized phase transitions \cite{Shi2016PRX}, Cooper pairs in superconductivity \cite{Bardeen1957PR}, and polarons in electron transport \cite{Holstein1959AP}. On the other hand, we notice that the hybrid system including few atoms and the coupled cavity array has been extensively studied in the past decade. It can be used for single-photon switch \cite{Shen2005PRL,chang2007NP}, quantum transistors \cite{zhou2008PRL}, routers \cite{zhou2013PRL}, supercavity \cite{zhou2008PRA}, non-reciprocal optics \cite{yao2023adp}, and frequency converters \cite{Wang2014PRA}. Inspired by these discoveries, in this paper, we propose a scheme to simulate the open quantum dynamics of an atom in a coupled resonators. Here, the coupled resonators form a structured bath with the energy band centered at the cavity frequency and bandwidth 4 times the inter-cavity coupling. When the atomic transition frequency lies within the energy band of the bath, the probability of the atomic excited state will appear to be periodic oscillations with a single frequency, i.e., regular oscillations, indicating that the hybrid system making up of the atom and the coupled resonators is in a bound state at this time. Therein, the bound state is the superposition of the atomic excited state and photons localized in a few cavities around the cavity where the atom exists.  The duration of the regular oscillations is finite and determined by the number of cavities in the coupled-cavity array. In this case, we find that the uncertainty of the measured transition frequency of the atom decreases over coding time as $t^{-1}$ in the duration of the regular oscillations. 

This paper is organized as follows. In the next section, we will introduce our simulation scheme. By using $N$-qubits, we can effectively simulate the quantum dynamics of one atom and $2^N-1$ coupled cavities. We also provide the probability of the atomic excited state and the standard deviation of the measured atomic transition frequency, of which the detailed derivations are respectively shown in Appendixes~\ref{AppendixA},\ref{AppendixB}.
Then, in Sec.~\ref{sec:Discussion}, we use the direct mapping method, which is given in Appendix~\ref{AppenixC}, to numerically simulate the dynamics of the atomic excited state.
 We explore the relation between the duration, frequency, amplitude, and mean of the regular oscillations and the number of cavities.  We use both analytical and numerical methods to verify that the error of the measurement decreases over time in the presence of a bound state. We find that the standard deviation of the measured atomic transition frequency on the scale $t^{-1}$ can be achieved. In Sec.~\ref{sec:Conclusion}, we summarize our main findings.


\section{Scheme}\label{Sec:Scheme}

\begin{figure}[htbp]
\centering
\includegraphics[width=8.6cm]{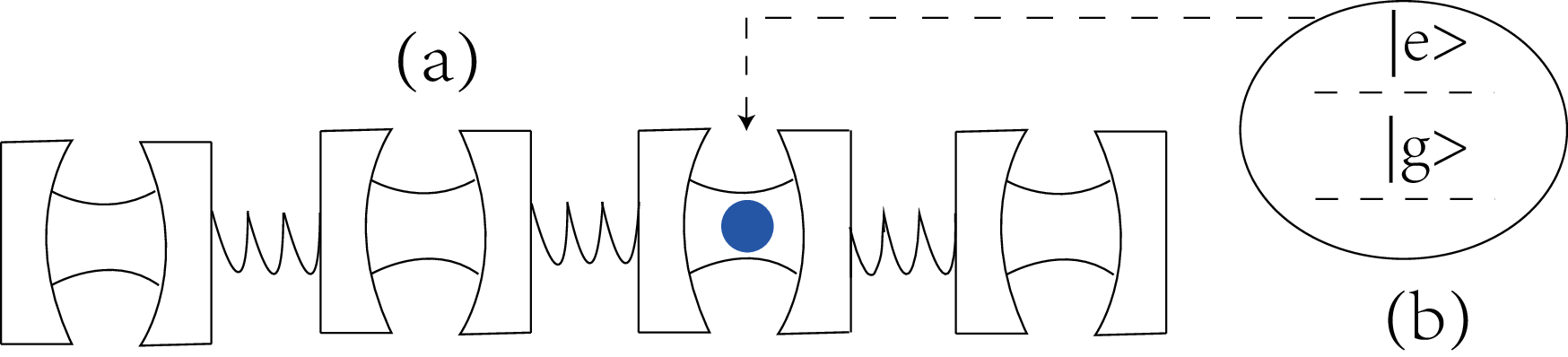}
\caption{Schematic illustration of a two-level atom and a coupled-cavity array. (a) Structure of coupled cavities, (b)a two-level atom. \label{fig:scheme}}
\end{figure}

We consider the interaction between a two-level atom and a coupled-cavity array \cite{bliokh2008PRM}. As shown in Fig.~\ref{fig:scheme}, the Hamiltonian of the system reads
\begin{eqnarray}
H&=&H_0+H_I,\label{eq:H}\\
H_0&\!=\!&\Omega|{e}\rangle\langle{e}|+\sum_{j=-j_M}^{j_M}[\omega_0a_{j}^{\dagger}a_{j}-\xi(a_{j}^{\dagger}a_{j+1}+a_{j+1}^{\dagger}a_{j})],\label{eq:ho}\\
H_I&=&J(a_{0}^{\dagger}\sigma_{-}+a_{0}\sigma_{+}),\label{eq:hi}
\end{eqnarray}
where $\Omega$ is the transition frequency of the atom, $a_{j}^{\dagger}$ ($a_{j}$) creates (annihilates) a photon with frequency $\omega_0$ in the $j$th cavity, $\xi$
is the coupling constant between two adjacent resonators, $\sigma_{+}=|{e}\rangle\langle{g}|=(\sigma_{-})^\dagger$ is the raising operator of the atom with $|{e}\rangle$ and $|{g}\rangle$ being the excited and ground state of the atom respectively, $J$ is the coupling constant between the atom and the $0$th cavity, $n$ is the total number of cavities, $j_M=(n-1)/2$. In this article, we have assumed $\hbar=1$ for simplicity.

Assuming the periodical boundary condition, by Fourier transformation, i.e., $a_j^{\dagger}=\sum_{k}a_k^{\dagger}e^{ikj}/\sqrt{n}$, the total Hamiltonian~(\ref{eq:H}) can be rewritten as
\begin{eqnarray}
H=\sum_{k}\omega_ka_k^{\dagger}a_k+\Omega|e\rangle\langle{e}|+\sum_{k}J_k(a_k^{\dagger}\sigma_{-}+\textrm{H.c.}),\label{eq:Hk}
\end{eqnarray}
where $\omega_k=\omega_0-2\xi\cos(k)$, $J_k=J/\sqrt{n}$. Since the total excitation $\mathcal{N}=\sum_ka^{\dagger}_ka_k+|{e}\rangle\langle{e}|$ is conserved, i.e., $[H,\mathcal{N}]=0$, we assume that at time $t$ the system is in the state $\vert\psi(t)\rangle=\alpha(t)|v,e\rangle+\sum_{k}\beta_{k}(t)a^{\dagger}_k|v,g\rangle$ with $|v\rangle$ being the vacuum state of all cavities. According to the Schr\"{o}dinger equation, we have
\begin{eqnarray}
i\dot{\alpha}&=&\Omega\alpha+\sum_{k}J_{k}\beta_{k},\label{eq:alpha}\\
i\dot{\beta}_{k}&=&\omega_{k}\beta_{k}+J_{k}\alpha,\label{eq:betak}
\end{eqnarray}
where the initial condition is given as $\alpha(0)=1,~\beta_{k}(0)=0$.
As shown in Appendix~\ref{AppendixA}, we can obtain the analytic expression of the probability amplitude of the excited state of the atom as
\begin{eqnarray}
\alpha(t)&=&A_{1}e^{p_{1}t}+A_{2}e^{p_{2}t}+\int_{-2\xi}^{2\xi}C(x)e^{i(x-\omega_{0})t}dx,\label{eq:alphat}
\end{eqnarray}
where $A_{j}$'s ($j=1,2$) are real, $p_j$'s ($j=1,2$) are imaginary, The explicit expressions of all these parameters including $C(x)$ can be referred to Appendix~\ref{AppendixA}.
Since the last term of Eq.~(\ref{eq:alphat}) decays exponentially in the long-time limit, the probability of the excited state of the atom reads
\begin{eqnarray}
P_e=|\alpha(\infty)|^{2}&=&A_{1}^{2}+A_{2}^{2}+2A_{1}A_{2}\cos(\phi t),\label{eq:infinity}
\end{eqnarray}
where $\phi=ip_{2}-ip_{1}$.
In order to measure the atomic transition frequency precisely, the variance is determined by the quantum Fisher information $F(\Omega)$ as \cite{he2021PRA,breuer2002theory,Liu2017PRA}
\begin{eqnarray}
\delta\Omega^{2}&=&\frac{1}{(T/t)F(\Omega)},\label{eq:deltaomega}\\
F(\Omega)&\equiv&\frac{1}{P_e(1-P_e)}\left(\frac{\partial P_e}{\partial\Omega}\right)^{2},\label{eq:Fisher}
\end{eqnarray}
where $T$ is the total duration of the experiment which is separated into repetition with each duration $t$. By the perturbation theory as shown in Appendix~\ref{AppendixB}, we have
\begin{eqnarray}
\phi=4\text{\ensuremath{\xi}}+\frac{(\omega_{0}-\Omega)^{2}+4\xi^{2}}{2\xi[(\omega_{0}-\Omega)^{2}-4\xi^{2}]^{2}}J^{4}.\label{eq:phi}
\end{eqnarray}
By substituting Eqs.~(\ref{eq:infinity}), (\ref{eq:Fisher}), (\ref{eq:phi}) into Eq.~(\ref{eq:deltaomega}), we can obtain the  uncertainty of the atomic transition frequency as
\begin{eqnarray}
\delta\Omega^{2}&=\frac{t\xi^2 B_1B_2}{J^{8}TB_3},\label{eq:uncertainty}
\end{eqnarray}
where
\begin{eqnarray}
B_1&=&-\frac{J^{8}}{4\Omega_{-}^{6}\xi^2}-\frac{J^{8}}{4\Omega_{+}^{6}\xi^2}+\frac{J^{8}\cos(\phi t)}{2\Omega_{-}^{3}\Omega_{+}^{3}\xi^2}+1,\\
B_2&=&\frac{1}{\Omega_{-}^{6}}+\frac{1}{(\Omega_{+})^{6}}-\frac{2\cos( \phi t)}{\Omega_{-}^{3}\Omega_{+}^{3}},\\
B_3^{1/2}&=&\frac{J^{4}t(\Omega-\omega_0)(12\xi^{2}+(\Omega-\omega_0)^{2})\sin(\phi t)}{\xi\Omega_{-}^{6}\Omega_{+}^{6}}\nonumber\\
&&-\frac{3}{\Omega_{-}^{7}}-\frac{3}{\Omega_{+}^{7}}+\frac{3\cos(\phi t)}{\Omega_{-}^{4}\Omega_{+}^{3}}+\frac{3\cos(\phi t)}{\Omega_{-}^{3}\Omega_{+}^{4}},
\end{eqnarray}
with $\Omega_{\pm}=\pm2\xi+\omega_{0}-\Omega$.

\section{Numerical Simulation and Discussion}
\label{sec:Discussion}

\begin{figure}[htbp]
\centering
\includegraphics[bb=00 0 380 300,width=8.5cm]
{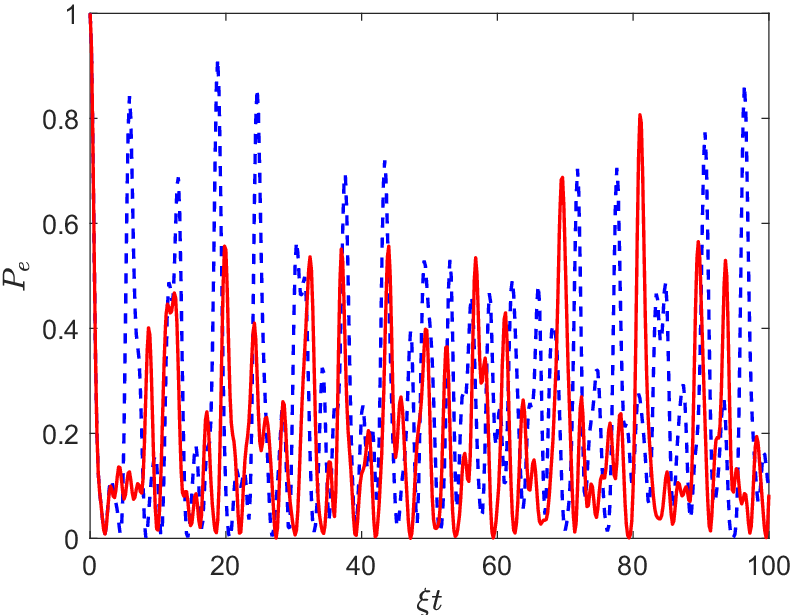}
\caption{The population dynamics $P_e(t)$ of atomic excited state with small $N$'s, e.g. $N=3$ (blue dashed line) and $N=4$ (red solid line). Here, the parameters used in the calculation are $\omega_0=10\xi$, $\Omega=11\xi$, and $J=1.3\xi$.}\label{fig:2ce12}
\end{figure}


\begin{figure}[htbp]
\centering
\includegraphics[bb=00 00 360 270,width=8.5cm]
{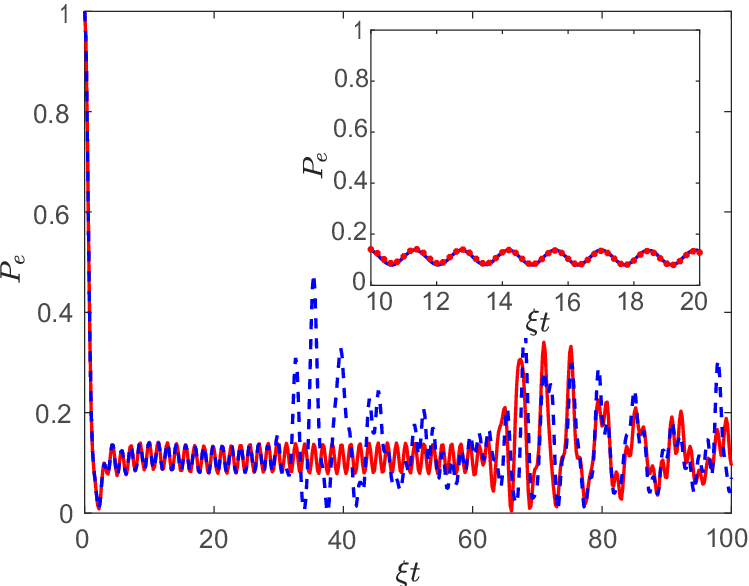}
\caption{The population dynamics $P_e(t)$ of atomic excited state with larger $N$'s, e.g. $N=6$ (blue dashed line) and $N=7$ (red solid line). The inset compares the numerical simulation of the regular oscillations of population dynamics $P_e(t)$ with $N=8$ (red stars) vs the analytical solution (blue solid line) for Eq.~(\ref{eq:infinity}) over a period of time. Here, the parameters are the same as Fig.~\ref{fig:2ce12}.}\label{fig:2ce13}
\end{figure}

As shown in Appendix~\ref{AppenixC}, we utilize $N$ qubits to effectively simulate the quantum dynamics for one atom interacting with $2^{N}-1$ coupled resonators in the single-excitation subspace.
In Fig.~\ref{fig:2ce12}, it is shown that when the number of qubits $N$ is relatively small, e.g. $N=3,4$, the population of the atom at the excited state $P_e(t)$ does not oscillate with a specific frequency, and thus does not provide any information for the frequency of the atom. However, if $N$ is increased, e.g. $N=6,7$ in Fig.~\ref{fig:2ce13}, we find that there are persistent oscillations in $P_e(t)$, which indicates that there exists a bound state in the atom interacting with the cavity array. The underlying physical mechanism can be well explained by the analytical solution obtained in the previous section, note that the analytical solution we obtained in the previous section is obtained under the assumption of infinite $N$. At the beginning, due to the third term in Eq.~(\ref{eq:alphat}), $P_e(t)$ oscillates out of order. As the time passes by, since the contribution from the branch cut vanishes, $P_e(t)$ oscillates with a specific frequency, as predicted by Eq.~(\ref{eq:infinity}). In the inset of Fig.~\ref{fig:2ce13}, the numerical simulation and Eq.~(\ref{eq:infinity})  fitting results are compared. The results of Fig.~\ref{fig:2ce13} show that when $N$ is large enough, e.g. $N=5$, the population of the excited state of the atom will show a regular oscillation for a period of time. The frequency, the amplitude, and the mean of these regular oscillations are respectively $\phi$, $2A_1A_2$, and $A_1^2+A_2^2$  in Eq.~(\ref{eq:infinity}). According to Eq.~(\ref{eq:phi}), $\phi$ is proportional to $J^4$. Contrary to $\phi$ vs $J$, the relation between $\phi$ and $\xi$ is more tricky. For a given $J$, there exists a minimum for $\phi$ as $\xi$ varies. As $J$ decreases, the minimum also decreases and the position of the minimum along the $\xi$-axis approaches closer and closer to the $J$-axis. $A_1^2+A_2^2$ is insensitive to the variation of $\xi$ and $J$ except when the atom is strongly coupled to the cavity and the cavities are weakly coupled to each other. In this regime, $A_1^2+A_2^2$ increases sharply as $J$ enlarges or $\xi$ decreases. The same behaviour can be also observed in the relation between $2A_1A_2$ and $J$ plus $\xi$. By measuring this frequency of regular oscillations, we effectively obtain the atomic transition frequency. Notice that these regular oscillations do not last for ever. Interestingly, the duration of these regular oscillations seems to get longer as $N$ increases. 


\begin{figure}[htbp]
\centering
\includegraphics[bb=0 0 365 290,width=8.5cm]
 {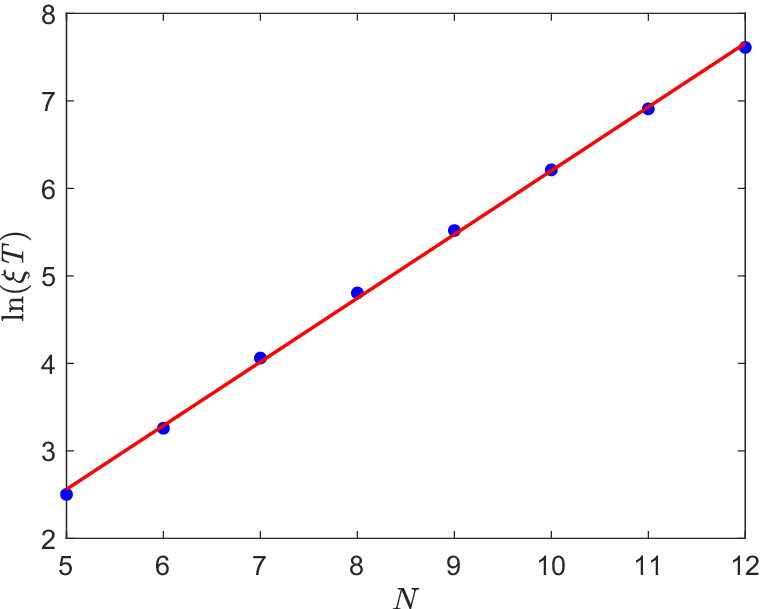}
\caption{The duration of the regular oscillations as a function of the number of qubits $N$. We use the function $\ln(\xi T)=0.72823\times N-1.0812$ (red solid line) to fit the data (blue dots) with the linear correlation coefficient $r=0.99967$. The other parameters are the same as Fig.~\ref{fig:2ce12}.}
\label{fig:fit}
\end{figure}

\begin{figure}[htbp]
\centering
\includegraphics[bb=00 0 370 290,width=8.5cm]
 {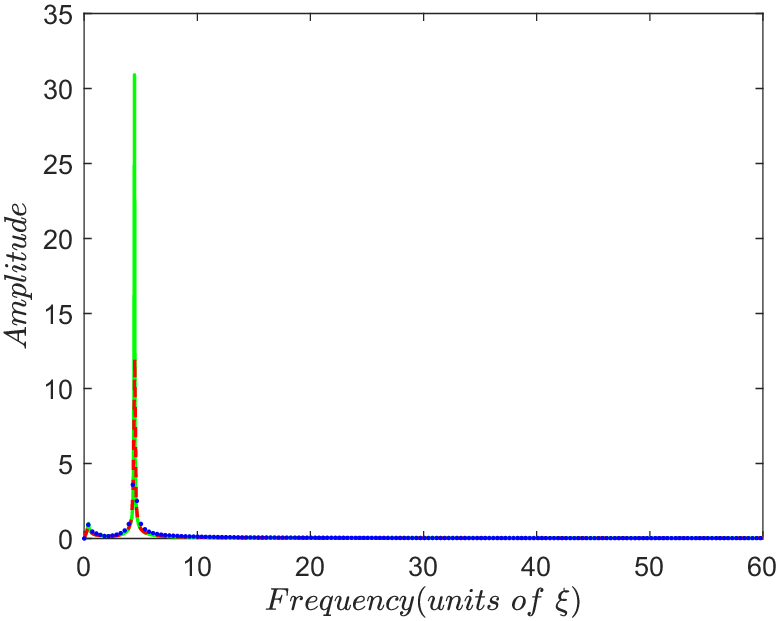}
\caption{Fourier transform for the regular oscillations with different qubit numbers. The FWHM of the main peak are respectively $\textrm{FWHM}=0.288\xi, 0.135\xi, 0.082\xi$ for $N=6$ by blue dotted line, $N=7$ by red dashed line, $N=8$ by green solid line. The other parameters are the same as Fig.~\ref{fig:2ce12}.}
\label{fig:fourier}
\end{figure}

\begin{figure}[htbp]
\centering
\includegraphics[bb=0 0 265 250,width=8.5cm]
 {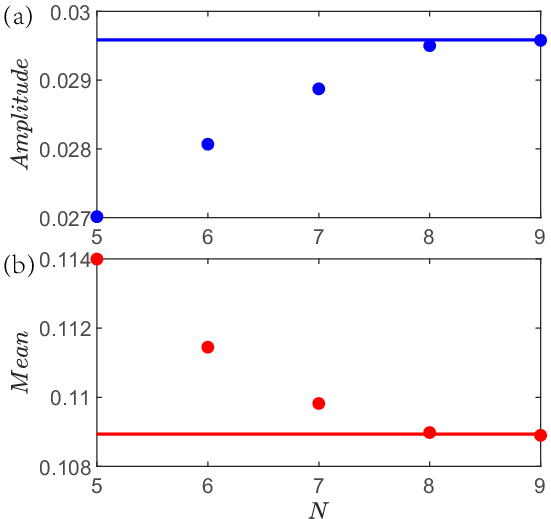}
\caption{(a) The amplitude of the regular oscillations for different numbers of qubits $N$'s by the numerical simulation (blue dots) is compared to the one with infinite $N$ obtained by the analytical solution (blue solid line). (b) The mean of the regular oscillations for different numbers of qubits $N$'s by the numerical simulation (red dots) is compared to the one with infinite $N$ obtained by the analytical solution (red solid line). The other parameters are the same as Fig.~\ref{fig:2ce12}.}
\label{fig:figF10_N56789}
\end{figure}


In order to study the relation between the duration of regular oscillations $T$ and the number of qubits $N$, we plot $\log(\xi T)$ vs $N$ in Fig.~\ref{fig:fit}. By a linear fitting with the linear
correlation coefficient $r=0.99$, we show that $\xi T=\exp(0.72823\times N-1.0812)$, which implies that the duration of the regular oscillations doubles as we use one more qubit, i.e., $\exp(0.72823)\simeq2.0714$. Thus, we can tune the coding time by varying the number of cavities in the array. It allows us to improve the measurement accuracy, so as to obtain the atomic transition frequency with a much smaller error. When $N$ approaches infinity, the coding time can be infinite, and we can obtain a perfect measurement of the atomic transition frequency, cf. Eq.~(\ref{eq:deltaomega}). To study their more properties, we perform the Fourier transform on the regular oscillations, as shown in Fig.~\ref{fig:fourier}. There is a profound main peak at $4.42\xi$. We can see that as the numbers of qubits $N$ increases, the height of the main peak is significantly enlarged and its full width at half maximum (FWHM) is reduced. It implies that the regular oscillations become more regular when $N$ approaches infinity.
We also notice that in addition to the main peak, there is a much smaller peak at $0.35\xi$. These suggest that the bound state is made up of atomic excitation and photons in a few cavities.
In Fig.~\ref{fig:figF10_N56789}, we show both the amplitude and mean of the regular oscillations. As the number of qubit $N$ increases, both the amplitude and mean of the regular oscillations approach the ones with infinite $N$ obtained by the analytical solution given in Eq.~(\ref{eq:infinity}). The results shown in Figs.~\ref{fig:fourier}-\ref{fig:figF10_N56789} suggest that the more qubits the system is composed of, the bound state appears more stable, and thus the less error we obtain when we measure the transition frequency of the atom. As shown in the figures below, we apply both the analytical and numerical methods to confirm our conjecture that bound states can help us improve the accuracy of the measurements.




\begin{figure}[htbp]
\centering
\includegraphics[bb=0 0 378 290,width=8.5cm]
 {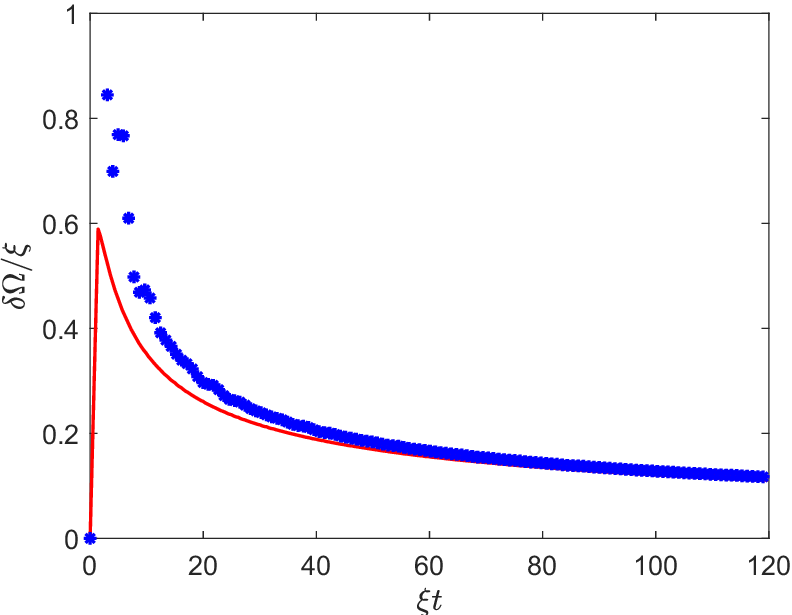}
\caption{Comparison of analytical and numerical results for the uncertainty $\delta\Omega$ vs time. The red solid line is obtained by Eq.~(\ref{eq:uncertainty}) while the blue stars are obtained by the numerically-exact solution with the following parameters, i.e., $N=8$, $\omega_0=20\xi,\Omega=20.5\xi$, $J=3\xi$. With the selection of $N$, the regular oscillations for the probability of the atomic excited state stops  at $\xi T=120$.}
\label{fig:uncertainty4}
\end{figure}


\begin{figure}[htbp]
\centering
\includegraphics[bb=0 0 380 290,width=8.5cm]
 {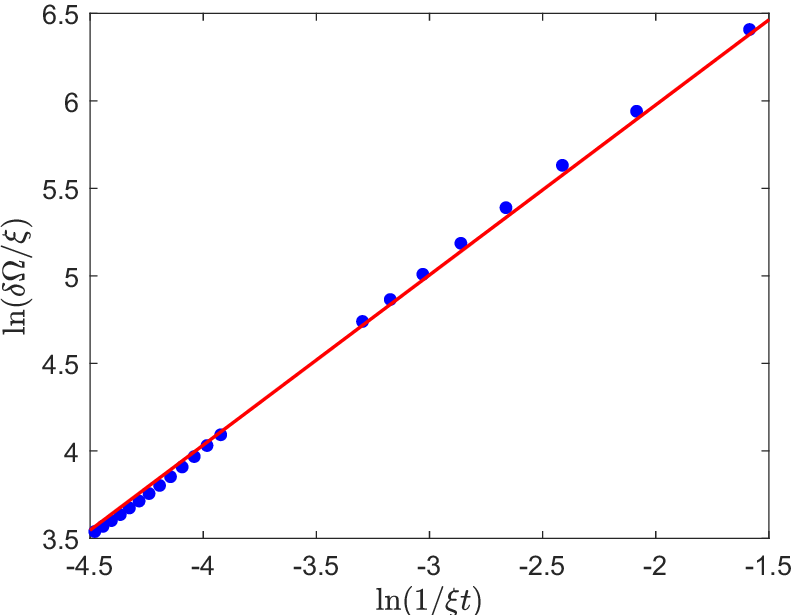}
\caption{The scaling of the uncertainty $\delta\Omega$ vs time. The blue dots are obtained by the numerical solution with the following parameters, i.e., $N=8$, $\omega_0=20\xi$, $\Omega=20\xi$, $J=0.3\xi$. The red solid line is obtained by fitting the data with the function $\textrm{ln}(\delta\Omega/\xi) =0.97196\times\textrm{ln}(1/{\xi t})+7.9211$ and the linear correlation coefficient $r=0.9995$.}
\label{fig:detaomega_t_sub2}
\end{figure}

\begin{figure}[htbp]
\centering
\includegraphics[bb=00 0 380 290,width=8.5cm]
 {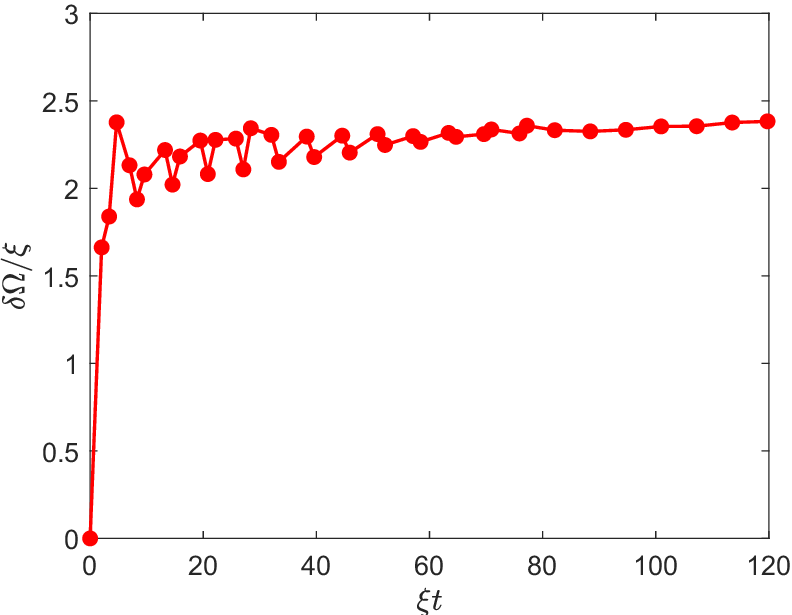}
\caption{Numerical simulation of the uncertainty $\delta\Omega$ vs time with the measured atomic transition frequency $\Omega$ beyond the band of the coupled resonators. The data is obtained by Eq.~(\ref{eq:deltaomega}) with the following parameters, i.e., $N=8$, $\Omega=17\xi$, $\omega_0=10\xi$, $J=0.3\xi$, and $\xi T=120$.}
\label{fig:shizhi2}
\end{figure}

 In Fig.~\ref{fig:uncertainty4}, by the red solid line, we show the uncertainty of atomic transition frequency $\delta\Omega$ obtained by Eq.~(\ref{eq:uncertainty}) when the atomic transition frequency is within the band of the coupled resonators. At first, $\delta\Omega$ experiences a rapid rise and  followed a slow decay, guaranteeing the recovery of vanishing measurement error. Notice that in obtaining Eq.~(\ref{eq:uncertainty}) we assume that the atom-resonator coupling $J$ is weak, i.e., $|\omega_0\pm2\xi|,\Omega\gg J$, and the atomic transition frequency is within the band of the coupled resonators, i.e., $\Omega\in[\omega_0-2\xi,\omega_0+2\xi]$. In order to verify that the above assumption is valid, we compare the approximated Eq.~(\ref{eq:uncertainty}) with numerically-exact Eq.~(\ref{eq:deltaomega}) in Fig.~\ref{fig:uncertainty4}. Obviously, both two solutions agree very well with each other and thus suggests reliable approximations made in Eq.~(\ref{eq:uncertainty}).  In a noiseless metrology, the standard deviation of the measured transition frequency scales as $t^{-1}$ with respect to the time. In Fig.~\ref{fig:detaomega_t_sub2}, we investigate the standard deviation at the optimal measurement times. By linear fitting, we numerically demonstrate that $\delta\Omega\propto t^{-1}$ up to $\xi T=120$, which is the duration of the regular oscillations.
However, due to the limitation of the approximations of the analytical method, we can not effectively simulate the situation with the atomic transition frequency beyond the band by Eq.~(\ref{eq:uncertainty}). Here, in Fig.~\ref{fig:shizhi2}, we plot the dynamics of the measured atomic transition frequency by Eq.~(\ref{eq:deltaomega}) with the parameters being $N=8$, $\Omega=17\xi$, $\omega_0=10\xi$, $J=0.3\xi$, and $\xi T=120$. As shown, after an increase about $\xi t=50$, $\delta\Omega$ achieves a steady state although there still exist oscillations with small amplitude. It does not recover the $1/t$ measurement because there does not exist the bound state.

%

\section{Conclusion}\label{sec:Conclusion}

To conclude, we study the effects of the bound state on the quantum metrology in an atom interacting with coupled resonators. In the hybrid system with finite number $n$ of cavities, we find that when the number of qubits is large enough, e.g. $n\geq2^5-1=31$, there exist regular oscillations in the population dynamics of the atomic excited state, which may be used for the evaluation of the atomic transition frequency. The duration of these regular oscillations is linear with respect to $n$. For a given $n$, during the duration of the regular oscillations, we show that the standard deviation of the measured atomic transition frequency is inversely proportional to $t$. If more cavities are involved, since the duration of the regular oscillations gets longer, the measured atomic transition frequency will be more and more accurate.
These regular oscillations indicate that there exists a bound state in the hybrid system. Their duration lasts for a finite time because there are finite cavities, i.e., finite bath modes. By numerical simulations, we can prove that the bound state exists when the atomic transition frequency is within the energy band of the coupled resonators.
In addition to the numerical simulations, we obtain an analytical result of the uncertainty of the atomic transition frequency for small atom-cavity interaction strength when the bound state is present. It indicates that the error-free measurement is recovered with infinite cavities in the long-time limit due to the existence of the bound state.
The results suggest that the non-Markovian effects and the bound state account for the high accuracy of the quantum metrology. Our research may shed light on the design of the hybrid system exploring the bound state for quantum metrology.

\begin{acknowledgments}

This work is supported by Innovation Program for Quantum Science and Technology under Grant No. 2023ZD0300200, Beijing Natural Science Foundation under Grant No.~1202017 and the National Natural Science Foundation of China under Grant Nos.~11674033,~11505007, and Beijing Normal University under Grant No.~2022129, Scientific and Technological Research Program of Chongqing Municipal Education Commission under Grant No.~KJQN202200603, and Chongqing University of Posts and Telecommunications under Grant No.~A2022-304.

\end{acknowledgments}

\appendix

\section{Population of Atomic Excited State}
\label{AppendixA}

After the Fourier transformation, the total Hamiltonian reads \cite{Ai2010CTP}
\begin{eqnarray}
H\!\!&=&\!\!\sum_{k}\omega_{k}a_{k}^{\dagger}a_{k}+\Omega|e\rangle\langle e|+\sum_{k}J_{k}a_{k}^{\dagger}\sigma_{-}+\rm{H.c.}\label{eq:A1}
\end{eqnarray}
In the subspace of single excitation, we assume the wavefunction as
\begin{eqnarray}
\vert\psi\rangle&=&\alpha(t)|0,e\rangle+\sum_{k}\beta_{k}(t)|k,g\rangle.\label{eq:A2}
\end{eqnarray}
By substituting Eqs.~(\ref{eq:A2}) and (\ref{eq:A1}) into the Schr\"{o}dinger equation, we can obtain
\begin{eqnarray}
i\dot{\alpha}(t)&=&\Omega\alpha(t)+\sum_{k}J_{k}\beta_{k}(t),\label{eq:A3}\\
i\dot{\beta_{k}}(t)&=&\omega_{k}\beta_{k}(t)+J_{k}\alpha(t),\label{eq:A4}
\end{eqnarray}
where $\alpha(0)=1$, $\beta_{k}(0)=0$ are the initial condition of the system. Then, we perform the Laplace transform on Eqs.~(\ref{eq:A3}) and (\ref{eq:A4}) to have
\begin{eqnarray}
i(p\widetilde{\alpha}(p)-1)&=&\Omega\widetilde{\alpha}(p)+\sum_{k}{J_{k}}\widetilde{\beta}_{k}(p),\label{eq:A5}\\
ip\widetilde{\beta}_{k}(p)&=&\omega_{k}\widetilde{\beta}_{k}(p)+J_{k}\widetilde{\alpha}(p).\label{eq:A6}
\end{eqnarray}
After some algebra, we can calculate the probability amplitude of the excited state in $p$-space as
\begin{eqnarray}
\widetilde{\alpha}(p)&=\frac{1}{p+i\Omega+\sum_{k}\frac{J_{k}^{2}}{p+i\omega_{k}}}.\label{eq:A7}
\end{eqnarray}
After the inverse Laplace transformation, we can obtain the probability amplitude of the excited state in the time domain as
\begin{eqnarray}
\alpha(t)&=&\frac{1}{2\pi i}\int_{\sigma-i\infty}^{\sigma+i\infty}dp\widetilde{\alpha}(p)e^{pt}.\label{eq:A8}
\end{eqnarray}
Using the residual theorem, we have
\begin{eqnarray}
\alpha(t)&=&-\frac{1}{2\pi i}(\int_{C_{R}}\widetilde{\alpha}(p)e^{pt}+\int_{l_{1}}\widetilde{\alpha}(p)e^{pt}
+\int_{l_{2}}\widetilde{\alpha}(p)e^{pt})\nonumber\\& &+\sum_{j=1}^{2}\textrm{res}(\widetilde{\alpha}(p_{j})e^{p_{j}t})
,\label{eq:A9}
\end{eqnarray}
where $C_{R}$ is the large semicircle at the infinite, $p_j$'s and $l_{j}$'s ($j=1,2$) are respectively the two singularities and two branch lines which are given by
\begin{eqnarray}
p+i\Omega+\sum_{k}\frac{J_{k}^{2}}{p+i\omega_{k}}&=&0.\label{eq:A10}
\end{eqnarray}
According to Eq.~(\ref{eq:A10}), the branch lines are defined by  $p+i\omega_{k}=0$, i.e., $p\in[ip_{m},ip_{M}]$, where $p_{m}=-\omega_{0}-2\xi$, $p_{M}=-\omega_{0}+2\xi$. Notice that
\begin{eqnarray}
\sum_{k}\frac{J_{k}^{2}}{p+i\omega_{k}}&=&\frac{J^{2}}{2\pi\xi}\oint_{|z|=1} dz\frac{1}{z^{2}+\mathcal{M}z+1},\label{eq:A11}
\end{eqnarray}
where $\mathcal{M}=(ip-\omega_{0})/\xi$, $z_{\pm}=(-\mathcal{M}\pm\sqrt{\mathcal{M}^{2}-4})/2$ are the two singularities. In other words, $ p\notin[-i(\omega_{0}+2\xi),-i(\omega_{0}-2\xi)]$, i.e., $\mathcal{M}>2$ or $\mathcal{M}<-2$. In the former case, since $ip>\omega_{0}+2\xi$, $-1<z_{+}<0$, $z_{-}<-1$, we have
\begin{eqnarray}
\frac{J^{2}}{2\pi\xi}\oint_{|z|=1}dz\frac{1}{z^{2}+\mathcal{M}z+1}=\frac{iJ^{2}}{\xi\sqrt{\mathcal{M}^{2}-4}}.\label{eq:A12}
\end{eqnarray}
By substituting Eq.~(\ref{eq:A12}) into Eq.~(\ref{eq:A10}), we can obtain
\begin{eqnarray}
p+i\Omega+i\frac{J^{2}}{\xi\sqrt{\mathcal{M}^{2}-4}}=f_{1}(p).\label{eq:f1}
\end{eqnarray}
Through $f_{1}(p_1)=0$ and $ip_1>\omega_{0}+2\xi$, we attain the first singularity. In the same way, for the case of $\mathcal{M}<-2$, since
\begin{eqnarray}
\frac{J^{2}}{2\pi\xi}\oint_{|z|=1}\frac{dz}{z^{2}+\mathcal{M}z+1}&=&-\frac{iJ^{2}}{\xi\sqrt{\mathcal{M}^{2}-4}},\label{eq:A13}
\end{eqnarray}
we have
\begin{eqnarray}
p+i\Omega-i\frac{J^{2}}{\xi\sqrt{\mathcal{M}^{2}-4}}=f_{2}(p).\label{eq:f2}
\end{eqnarray}
By $f_{2}(p_2)=0$ and $ip_2<\omega_{0}-2\xi$, we can obtain the second singularity. As a result, the contribution from the singularities in Eq.~(\ref{eq:A9}) reads
\begin{eqnarray}
\textrm{res}\left(\widetilde{\alpha}(p_{j})e^{p_{j}t}\right)&=&A_{j}e^{p_{j}t},\label{eq:A14}
\end{eqnarray}
where $A_{j}=\left(\frac{df_{j}}{dp}|_{p_{j}}\right)^{-1}$.

Thanks to Jordan Lemma, we know $\int_{C_{R}}\widetilde{\alpha}(p)e^{pt}=0$. Thus, we have
\begin{align}
&-\frac{1}{2\pi i}(\int_{l_{1}}\widetilde{\alpha}(p)e^{pt}+\int_{l_{2}}\widetilde{\alpha}(p)e^{pt}dp) \nonumber\\
=&-\frac{1}{2\pi i}(\int_{p_{m}}^{p_{M}}\frac{e^{ipt}}{p+\Omega-\sum_{k}\frac{g_{k}^{2}}{p+\omega_{k}+i0^{+}}}dp\nonumber\\
&+\int_{p_{M}}^{p_{m}}\frac{e^{ipt}}{p+\Omega-\sum_{k}\frac{J_{k}^{2}}{p+\omega_{k}-i0^{+}}}dp).
\label{eq:A15}\end{align}
Here,
\begin{eqnarray}
&&\sum_{k}\frac{J_{k}^{2}}{p+i\omega_{k}\pm i0^{+}}\nonumber\\
&=&\frac{J^{2}}{2\pi}\left[\mp\frac{i2\pi}{\sqrt{4\xi^{2}-(\omega_{0}+p)^{2}}}+\int_{-\pi}^{\pi}dkP\left(\frac{1}{p+\omega_{k}}\right)\right],\nonumber\\
\label{eq:A16}\end{eqnarray}
where $P(x)$ is the principal value function.
Since
\begin{eqnarray}
\int_{-\pi}^{\pi}dkP\left(\frac{1}{p+\omega_{k}}\right)&=&0,\label{eq:A17}
\end{eqnarray}
by substituting Eq.~(\ref{eq:A16}) into Eq.~(\ref{eq:A15}), we have
\begin{eqnarray}
-\frac{1}{2\pi i}(\int_{l_{1}}\widetilde{\alpha}(p)e^{pt}+\int_{l_{2}}\widetilde{\alpha}(p)e^{pt}dp) &=&\int_{-2\xi}^{2\xi}C(x)e^{i(x-\omega_{0})t}dx,\nonumber\\
\label{eq:A18}
\end{eqnarray}
where $C(x)=(g^{2}\sqrt{4\xi^{2}-x^{2}})/[(\Omega-\omega_{0}+x)^{2}(4\xi^{2}-(x)^{2})+g^{4}]$.
Finally, we obtain the probability amplitude of the atomic excited state as
\begin{eqnarray}
\alpha(t)\!\!&=&\!\!A_{1}e^{p_{1}t}+A_{2}e^{p_{2}t}+\int_{-2\xi}^{2\xi}C(x)e^{i(x-\omega_{0})t}dx,\label{eq:A19}
\end{eqnarray}
where
\begin{eqnarray}
A_j&=&\frac{(ip_{j}-\omega_{0})^{2}-4\xi^{2}}{(ip_{j}-\omega_{0})^{2}-4\xi^{2}+(ip_{j}-\Omega)(ip_{j}-\omega_{0})}.\label{eq:A22}
\end{eqnarray}
After some algebra, we find that $p_j$'s ($j=1,2$) are pure imaginary numbers.

\section{Uncertainty by perturbation}
\label{AppendixB}

In order to obtain the uncertainty of $\Omega$, we shall first of all determine $p_1$ and $p_2$ in Eq.~(\ref{eq:A19}) by perturbation theory. Take advantage of Eqs.~(\ref{eq:f1}) and (\ref{eq:f2}), we have
\begin{eqnarray}
(ip-\Omega)^{2}[(ip-\omega_{0})^{2}-4\xi^{2}]&=&J^{4},\label{eq:B1}
\end{eqnarray}
By the perturbation theory, to the zeroth order of $J
^4$, we can obtain
\begin{eqnarray}
(x-\Omega)^{2}[(x-\omega_{0})^{2}-4\xi^{2}]&=&0,\label{eq:B2}
\end{eqnarray}
where $x=ip$. There are four solutions to the above equation, i.e.,
\begin{eqnarray}
x_{01}&=&\omega_{0}+2\xi,\label{eq:B3_1}\\
x_{02}&=&\omega_{0}-2\xi,\label{eq:B3_2}\\
x_{03}&=&x_{04}=\Omega\label{eq:B3_3}.
\end{eqnarray}

In the following, we will obtain the solutions to Eq.~(\ref{eq:B1}) to the first order of $J^4$.

As shown in Appendix~\ref{AppendixA}, i.e., $ip_{1}>2\xi+\omega_{0}$ and $ip_{2}<-2\xi+\omega_{0}$, only $x_{1}$ and $x_{2}$ are kept for the following discussions. To the first order of $J^4$, we assume $x_{j}=x_{0j}+C_{j}J^{4}$ ($j=1,2,3,4$).
Substitute them back into Eq.~(\ref{eq:B1}), we can obtain
\begin{eqnarray}
(x-\Omega)^{2}(x-x_{01})(x-x_{02})&=&J^{4}.\label{eq:B5}
\end{eqnarray}
Thus, we have
\begin{eqnarray}
C_{1}&=&\frac{1}{4\xi\Omega_{+}^{2}},\\
C_{2}&=&\frac{-1}{4\xi\Omega_{-}^{2}}.\label{eq:B7}
\end{eqnarray}
where $\Omega_{\pm}=\pm2\xi+\omega_{0}-\Omega$.
In all, $p_1=-ix_1$ and $p_2=-ix_2$.

Having known the analytical expressions of $p_j$'s, in the long-time limit, the population of the atom in the excited state reads
\begin{eqnarray}
P_e=|\alpha(\infty)|^{2}&=&A_{1}^{2}+A_{2}^{2}+2A_{1}A_{2}\cos(\phi t),\label{eq:B8}
\end{eqnarray}
where $\phi=x_1-x_2$, the contribution from the integral term in Eq.~(\ref{eq:A19}) vanishes when the time is large enough. By substituting $p_1$ and $p_2$ into Eq.~(\ref{eq:A22}), we have
\begin{eqnarray}
A_{1}&=&\frac{J^{4}}{2\Omega_{+}^{3}\xi},\label{eq:B9_1}\\
A_{2}&=&-\frac{J^{4}}{2\Omega_{-}^{3}\xi},\label{eq:B9_2}\\
\phi&=&4\xi+(C_{1}-C_{2})J^{4}.\label{eq:B9_3}
\end{eqnarray}

The uncertainty of the atomic transition frequency is written as
\begin{eqnarray}
\delta\Omega^{2}&=&\frac{1}{(T/t)F(\Omega)},\label{eq:B10}
\end{eqnarray}
where the Fisher information $F(\Omega)$ is
\begin{eqnarray}
F(\Omega)&\equiv&\frac{1}{P_e(1-P_e)}\left(\frac{\partial P_e}{\partial\Omega}\right)^{2}.\label{eq:B11}
\end{eqnarray}
According to Eqs.~(\ref{eq:B10}), (\ref{eq:B11}), (\ref{eq:B8}), the uncertainty is explicitly given as
\begin{eqnarray}
\delta\Omega^{2}&=\frac{t\xi^2 B_1B_2}{J^{8}TB_3},\label{eq:B12}
\end{eqnarray}
where
\begin{eqnarray}
B_1&=&-\frac{J^{8}}{4\Omega_{-}^{6}\xi^2}-\frac{J^{8}}{4\Omega_{+}^{6}\xi^2}+\frac{J^{8}\cos(\phi t)}{2\Omega_{-}^{3}\Omega_{+}^{3}\xi^2}+1,\\
B_2&=&\frac{1}{\Omega_{-}^{6}}+\frac{1}{(\Omega_{+})^{6}}-\frac{2\cos( \phi t)}{\Omega_{-}^{3}\Omega_{+}^{3}},\\
B_3^{1/2}&=&\frac{J^{4}t(\Omega-\omega_0)(12\xi^{2}+(\Omega-\omega_0)^{2})\sin(\phi t)}{\xi\Omega_{-}^{6}\Omega_{+}^{6}}\nonumber\\
&&-\frac{3}{\Omega_{-}^{7}}-\frac{3}{\Omega_{+}^{7}}+\frac{3\cos(\phi t)}{\Omega_{-}^{4}\Omega_{+}^{3}}+\frac{3\cos(\phi t)}{\Omega_{-}^{3}\Omega_{+}^{4}}.
\end{eqnarray}

\section{Direct Mapping}
\label{AppenixC}

In this section, we discuss how to demonstrate the bound-state enhanced metrology by the quantum simulation approach with finite number of  qubits.
We focus our investigation in the single-excitation subspace. If we use $N$ qubits for quantum simulation, the dimension of the Hilbert space is $2^N$. For example, when $N=2$, we perform the following mapping before quantum simulation, i.e.,
\begin{eqnarray}
|{00}\rangle&=&|0\rangle|{e}\rangle,\\
|{01}\rangle&=&a_1^\dagger|0\rangle|{g}\rangle,\\
|{10}\rangle&=&a_2^\dagger|0\rangle|{g}\rangle,\\
|{11}\rangle&=&a_3^\dagger|0\rangle|{g}\rangle,
\end{eqnarray}
where $|{g}\rangle$ ($|{e}\rangle$) the atomic ground (excited) state, $|0\rangle$ is the vacuum state of all cavities, $a_j^\dagger|0\rangle$ indicates that there is one photon in the $j$th cavity while all other cavities are in the vacuum state. Therefore, by using $N$ qubits, we can effectively simulation the quantum dynamics of one atom plus number $2^N-1$ coupled cavities.



\begin{thebibliography}{56}%
\makeatletter
\providecommand \@ifxundefined [1]{%
 \@ifx{#1\undefined}
}%
\providecommand \@ifnum [1]{%
 \ifnum #1\expandafter \@firstoftwo
 \else \expandafter \@secondoftwo
 \fi
}%
\providecommand \@ifx [1]{%
 \ifx #1\expandafter \@firstoftwo
 \else \expandafter \@secondoftwo
 \fi
}%
\providecommand \natexlab [1]{#1}%
\providecommand \enquote  [1]{``#1''}%
\providecommand \bibnamefont  [1]{#1}%
\providecommand \bibfnamefont [1]{#1}%
\providecommand \citenamefont [1]{#1}%
\providecommand \href@noop [0]{\@secondoftwo}%
\providecommand \href [0]{\begingroup \@sanitize@url \@href}%
\providecommand \@href[1]{\@@startlink{#1}\@@href}%
\providecommand \@@href[1]{\endgroup#1\@@endlink}%
\providecommand \@sanitize@url [0]{\catcode `\\12\catcode `\$12\catcode
  `\&12\catcode `\#12\catcode `\^12\catcode `\_12\catcode `\%12\relax}%
\providecommand \@@startlink[1]{}%
\providecommand \@@endlink[0]{}%
\providecommand \url  [0]{\begingroup\@sanitize@url \@url }%
\providecommand \@url [1]{\endgroup\@href {#1}{\urlprefix }}%
\providecommand \urlprefix  [0]{URL }%
\providecommand \Eprint [0]{\href }%
\providecommand \doibase [0]{http://dx.doi.org/}%
\providecommand \selectlanguage [0]{\@gobble}%
\providecommand \bibinfo  [0]{\@secondoftwo}%
\providecommand \bibfield  [0]{\@secondoftwo}%
\providecommand \translation [1]{[#1]}%
\providecommand \BibitemOpen [0]{}%
\providecommand \bibitemStop [0]{}%
\providecommand \bibitemNoStop [0]{.\EOS\space}%
\providecommand \EOS [0]{\spacefactor3000\relax}%
\providecommand \BibitemShut  [1]{\csname bibitem#1\endcsname}%
\let\auto@bib@innerbib\@empty

\bibitem [{\citenamefont {Grote}\ \emph {et~al.}(2013)\citenamefont {Grote},
  \citenamefont {Danzmann}, \citenamefont {Dooley}, \citenamefont {Schnabel},
  \citenamefont {Slutsky},\ and\ \citenamefont {Vahlbruch}}]{grote2013PRL}%
  \BibitemOpen
  \bibfield  {author} {\bibinfo {author} {\bibfnamefont {H.}~\bibnamefont
  {Grote}}, \bibinfo {author} {\bibfnamefont {K.}~\bibnamefont {Danzmann}},
  \bibinfo {author} {\bibfnamefont {K.~L.}\ \bibnamefont {Dooley}}, \bibinfo
  {author} {\bibfnamefont {R.}~\bibnamefont {Schnabel}}, \bibinfo {author}
  {\bibfnamefont {J.}~\bibnamefont {Slutsky}}, \ and\ \bibinfo {author}
  {\bibfnamefont {H.}~\bibnamefont {Vahlbruch}},\ }\bibfield  {title} {\enquote
  {\bibinfo {title} {First long-term application of squeezed states of light in
  a gravitational-wave observatory},}\ }\href {\doibase
  https://doi.org/10.1103/PhysRevLett.110.181101} {\bibfield  {journal}
  {\bibinfo  {journal} {Phys. Rev. Lett.}\ }\textbf {\bibinfo {volume} {110}},\
  \bibinfo {pages} {181101} (\bibinfo {year} {2013})}\BibitemShut {NoStop}%
\bibitem [{\citenamefont {Schnabel}\ \emph {et~al.}(2010)\citenamefont
  {Schnabel}, \citenamefont {Mavalvala}, \citenamefont {McClelland},\ and\
  \citenamefont {Lam}}]{schnabel2010NC}%
  \BibitemOpen
  \bibfield  {author} {\bibinfo {author} {\bibfnamefont {R.}~\bibnamefont
  {Schnabel}}, \bibinfo {author} {\bibfnamefont {N.}~\bibnamefont {Mavalvala}},
  \bibinfo {author} {\bibfnamefont {D.~E.}\ \bibnamefont {McClelland}}, \ and\
  \bibinfo {author} {\bibfnamefont {P.~K.}\ \bibnamefont {Lam}},\ }\bibfield
  {title} {\enquote {\bibinfo {title} {Quantum metrology for gravitational wave
  astronomy},}\ }\href {\doibase https://doi.org/10.1038/ncomms1122} {\bibfield
   {journal} {\bibinfo  {journal} {Nat. Commun.}\ }\textbf {\bibinfo {volume}
  {1}},\ \bibinfo {pages} {121} (\bibinfo {year} {2010})}\BibitemShut {NoStop}%
\bibitem [{\citenamefont {{The LIGO Scientific
  Collaboration}}(2011)}]{ligo2011NP}%
  \BibitemOpen
  \bibfield  {author} {\bibinfo {author} {\bibnamefont {{The LIGO Scientific
  Collaboration}}},\ }\bibfield  {title} {\enquote {\bibinfo {title} {A
  gravitational wave observatory operating beyond the quantum shot-noise
  limit},}\ }\href {\doibase https://doi.org/10.1038/nphys2083} {\bibfield
  {journal} {\bibinfo  {journal} {Nat. Phys.}\ }\textbf {\bibinfo {volume}
  {7}},\ \bibinfo {pages} {962} (\bibinfo {year} {2011})}\BibitemShut {NoStop}%
\bibitem [{\citenamefont {Barzanjeh}\ \emph {et~al.}(2015)\citenamefont
  {Barzanjeh}, \citenamefont {Guha}, \citenamefont {Weedbrook}, \citenamefont
  {Vitali}, \citenamefont {Shapiro},\ and\ \citenamefont
  {Pirandola}}]{barzanjeh2015PRL}%
  \BibitemOpen
  \bibfield  {author} {\bibinfo {author} {\bibfnamefont {S.}~\bibnamefont
  {Barzanjeh}}, \bibinfo {author} {\bibfnamefont {S.}~\bibnamefont {Guha}},
  \bibinfo {author} {\bibfnamefont {C.}~\bibnamefont {Weedbrook}}, \bibinfo
  {author} {\bibfnamefont {D.}~\bibnamefont {Vitali}}, \bibinfo {author}
  {\bibfnamefont {J.~H.}\ \bibnamefont {Shapiro}}, \ and\ \bibinfo {author}
  {\bibfnamefont {S.}~\bibnamefont {Pirandola}},\ }\bibfield  {title} {\enquote
  {\bibinfo {title} {Microwave quantum illumination},}\ }\href {\doibase
  https://doi.org/10.1103/PhysRevLett.114.080503} {\bibfield  {journal}
  {\bibinfo  {journal} {Phys. Rev. Lett.}\ }\textbf {\bibinfo {volume} {114}},\
  \bibinfo {pages} {080503} (\bibinfo {year} {2015})}\BibitemShut {NoStop}%
\bibitem [{\citenamefont {Maccone}\ and\ \citenamefont
  {Ren}(2020)}]{maccone2020PRL}%
  \BibitemOpen
  \bibfield  {author} {\bibinfo {author} {\bibfnamefont {L.}~\bibnamefont
  {Maccone}}\ and\ \bibinfo {author} {\bibfnamefont {C.~L.}\ \bibnamefont
  {Ren}},\ }\bibfield  {title} {\enquote {\bibinfo {title} {Quantum radar},}\
  }\href {\doibase https://doi.org/10.1103/PhysRevLett.124.200503} {\bibfield
  {journal} {\bibinfo  {journal} {Phys. Rev. Lett.}\ }\textbf {\bibinfo
  {volume} {124}},\ \bibinfo {pages} {200503} (\bibinfo {year}
  {2020})}\BibitemShut {NoStop}%
\bibitem [{\citenamefont {Arrad}\ \emph {et~al.}(2014)\citenamefont {Arrad},
  \citenamefont {Vinkler}, \citenamefont {Aharonov},\ and\ \citenamefont
  {Retzker}}]{arrad2014PRL}%
  \BibitemOpen
  \bibfield  {author} {\bibinfo {author} {\bibfnamefont {G.}~\bibnamefont
  {Arrad}}, \bibinfo {author} {\bibfnamefont {Y.}~\bibnamefont {Vinkler}},
  \bibinfo {author} {\bibfnamefont {D.}~\bibnamefont {Aharonov}}, \ and\
  \bibinfo {author} {\bibfnamefont {A.}~\bibnamefont {Retzker}},\ }\bibfield
  {title} {\enquote {\bibinfo {title} {Increasing sensing resolution with error
  correction},}\ }\href {\doibase
  https://doi.org/10.1103/PhysRevLett.112.150801} {\bibfield  {journal}
  {\bibinfo  {journal} {Phys. Rev. Lett.}\ }\textbf {\bibinfo {volume} {112}},\
  \bibinfo {pages} {150801} (\bibinfo {year} {2014})}\BibitemShut {NoStop}%
\bibitem [{\citenamefont {Kruse}\ \emph {et~al.}(2016)\citenamefont {Kruse},
  \citenamefont {Lange}, \citenamefont {Peise}, \citenamefont {L{\"u}cke},
  \citenamefont {Pezz{\`e}}, \citenamefont {Arlt}, \citenamefont {Ertmer},
  \citenamefont {Lisdat}, \citenamefont {Santos}, \citenamefont {Smerzi},\ and\
  \citenamefont {Klempt}}]{kruse2016PRL}%
  \BibitemOpen
  \bibfield  {author} {\bibinfo {author} {\bibfnamefont {I.}~\bibnamefont
  {Kruse}}, \bibinfo {author} {\bibfnamefont {K.}~\bibnamefont {Lange}},
  \bibinfo {author} {\bibfnamefont {J.}~\bibnamefont {Peise}}, \bibinfo
  {author} {\bibfnamefont {B.}~\bibnamefont {L{\"u}cke}}, \bibinfo {author}
  {\bibfnamefont {L.}~\bibnamefont {Pezz{\`e}}}, \bibinfo {author}
  {\bibfnamefont {J.}~\bibnamefont {Arlt}}, \bibinfo {author} {\bibfnamefont
  {W.}~\bibnamefont {Ertmer}}, \bibinfo {author} {\bibfnamefont
  {C.}~\bibnamefont {Lisdat}}, \bibinfo {author} {\bibfnamefont
  {L.}~\bibnamefont {Santos}}, \bibinfo {author} {\bibfnamefont
  {A.}~\bibnamefont {Smerzi}}, \ and\ \bibinfo {author} {\bibfnamefont
  {C.}~\bibnamefont {Klempt}},\ }\bibfield  {title} {\enquote {\bibinfo {title}
  {Improvement of an atomic clock using squeezed vacuum},}\ }\href {\doibase
  https://doi.org/10.1103/PhysRevLett.117.143004} {\bibfield  {journal}
  {\bibinfo  {journal} {Phys. Rev. Lett.}\ }\textbf {\bibinfo {volume} {117}},\
  \bibinfo {pages} {143004} (\bibinfo {year} {2016})}\BibitemShut {NoStop}%
\bibitem [{\citenamefont {Hosten}\ \emph {et~al.}(2016)\citenamefont {Hosten},
  \citenamefont {Engelsen}, \citenamefont {Krishnakumar},\ and\ \citenamefont
  {Kasevich}}]{hosten2016Nat}%
  \BibitemOpen
  \bibfield  {author} {\bibinfo {author} {\bibfnamefont {O.}~\bibnamefont
  {Hosten}}, \bibinfo {author} {\bibfnamefont {N.~J.}\ \bibnamefont
  {Engelsen}}, \bibinfo {author} {\bibfnamefont {R.}~\bibnamefont
  {Krishnakumar}}, \ and\ \bibinfo {author} {\bibfnamefont {M.~A.}\
  \bibnamefont {Kasevich}},\ }\bibfield  {title} {\enquote {\bibinfo {title}
  {Measurement noise 100 times lower than the quantum-projection limit using
  entangled atoms},}\ }\href {\doibase https://doi.org/10.1038/nature16176}
  {\bibfield  {journal} {\bibinfo  {journal} {Nature}\ }\textbf {\bibinfo
  {volume} {529}},\ \bibinfo {pages} {505} (\bibinfo {year}
  {2016})}\BibitemShut {NoStop}%
\bibitem [{\citenamefont {Pezz{\`e}}\ and\ \citenamefont
  {Smerzi}(2020)}]{pezze2020PRL}%
  \BibitemOpen
  \bibfield  {author} {\bibinfo {author} {\bibfnamefont {L.}~\bibnamefont
  {Pezz{\`e}}}\ and\ \bibinfo {author} {\bibfnamefont {A.}~\bibnamefont
  {Smerzi}},\ }\bibfield  {title} {\enquote {\bibinfo {title}
  {Heisenberg-limited noisy atomic clock using a hybrid coherent and squeezed
  state protocol},}\ }\href {\doibase
  https://doi.org/10.1103/PhysRevLett.125.210503} {\bibfield  {journal}
  {\bibinfo  {journal} {Phys. Rev. Lett.}\ }\textbf {\bibinfo {volume} {125}},\
  \bibinfo {pages} {210503} (\bibinfo {year} {2020})}\BibitemShut {NoStop}%
\bibitem [{\citenamefont {Kaubruegger}\ \emph {et~al.}(2021)\citenamefont
  {Kaubruegger}, \citenamefont {Vasilyev}, \citenamefont {Schulte},
  \citenamefont {Hammerer},\ and\ \citenamefont {Zoller}}]{kaubruegger2021PRX}%
  \BibitemOpen
  \bibfield  {author} {\bibinfo {author} {\bibfnamefont {R.}~\bibnamefont
  {Kaubruegger}}, \bibinfo {author} {\bibfnamefont {D.~V.}\ \bibnamefont
  {Vasilyev}}, \bibinfo {author} {\bibfnamefont {M.}~\bibnamefont {Schulte}},
  \bibinfo {author} {\bibfnamefont {K.}~\bibnamefont {Hammerer}}, \ and\
  \bibinfo {author} {\bibfnamefont {P.}~\bibnamefont {Zoller}},\ }\bibfield
  {title} {\enquote {\bibinfo {title} {Quantum variational optimization of
  ramsey interferometry and atomic clocks},}\ }\href {\doibase
  https://doi.org/10.1103/PhysRevX.11.041045} {\bibfield  {journal} {\bibinfo
  {journal} {Phys. Rev. X}\ }\textbf {\bibinfo {volume} {11}},\ \bibinfo
  {pages} {041045} (\bibinfo {year} {2021})}\BibitemShut {NoStop}%
\bibitem [{\citenamefont {Thiel}\ \emph {et~al.}(2016)\citenamefont {Thiel},
  \citenamefont {Rohner}, \citenamefont {Ganzhorn}, \citenamefont {Appel},
  \citenamefont {Neu}, \citenamefont {M{\"u}ller}, \citenamefont {Kleiner},
  \citenamefont {Koelle},\ and\ \citenamefont {Maletinsky}}]{thiel2016NN}%
  \BibitemOpen
  \bibfield  {author} {\bibinfo {author} {\bibfnamefont {L.}~\bibnamefont
  {Thiel}}, \bibinfo {author} {\bibfnamefont {D.}~\bibnamefont {Rohner}},
  \bibinfo {author} {\bibfnamefont {M.}~\bibnamefont {Ganzhorn}}, \bibinfo
  {author} {\bibfnamefont {P.}~\bibnamefont {Appel}}, \bibinfo {author}
  {\bibfnamefont {E.}~\bibnamefont {Neu}}, \bibinfo {author} {\bibfnamefont
  {B.}~\bibnamefont {M{\"u}ller}}, \bibinfo {author} {\bibfnamefont
  {R.}~\bibnamefont {Kleiner}}, \bibinfo {author} {\bibfnamefont
  {D.}~\bibnamefont {Koelle}}, \ and\ \bibinfo {author} {\bibfnamefont
  {P.}~\bibnamefont {Maletinsky}},\ }\bibfield  {title} {\enquote {\bibinfo
  {title} {Quantitative nanoscale vortex imaging using a cryogenic quantum
  magnetometer},}\ }\href {\doibase https://doi.org/10.1038/nnano.2016.63}
  {\bibfield  {journal} {\bibinfo  {journal} {Nat. Nanotechnol.}\ }\textbf
  {\bibinfo {volume} {11}},\ \bibinfo {pages} {677} (\bibinfo {year}
  {2016})}\BibitemShut {NoStop}%
\bibitem [{\citenamefont {Taylor}\ \emph {et~al.}(2008)\citenamefont {Taylor},
  \citenamefont {Cappellaro}, \citenamefont {Childress}, \citenamefont {Jiang},
  \citenamefont {Budker}, \citenamefont {Hemmer}, \citenamefont {Yacoby},
  \citenamefont {Walsworth},\ and\ \citenamefont {Lukin}}]{taylor2008NP}%
  \BibitemOpen
  \bibfield  {author} {\bibinfo {author} {\bibfnamefont {J.~M.}\ \bibnamefont
  {Taylor}}, \bibinfo {author} {\bibfnamefont {P.}~\bibnamefont {Cappellaro}},
  \bibinfo {author} {\bibfnamefont {L.}~\bibnamefont {Childress}}, \bibinfo
  {author} {\bibfnamefont {L.}~\bibnamefont {Jiang}}, \bibinfo {author}
  {\bibfnamefont {D.}~\bibnamefont {Budker}}, \bibinfo {author} {\bibfnamefont
  {P.~R.}\ \bibnamefont {Hemmer}}, \bibinfo {author} {\bibfnamefont
  {A.}~\bibnamefont {Yacoby}}, \bibinfo {author} {\bibfnamefont
  {R.}~\bibnamefont {Walsworth}}, \ and\ \bibinfo {author} {\bibfnamefont
  {M.~D.}\ \bibnamefont {Lukin}},\ }\bibfield  {title} {\enquote {\bibinfo
  {title} {High-sensitivity diamond magnetometer with nanoscale resolution},}\
  }\href {\doibase https://doi.org/10.1038/nphys1075} {\bibfield  {journal}
  {\bibinfo  {journal} {Nat. Phys.}\ }\textbf {\bibinfo {volume} {4}},\
  \bibinfo {pages} {810} (\bibinfo {year} {2008})}\BibitemShut {NoStop}%
\bibitem [{\citenamefont {Bao}\ \emph {et~al.}(2020)\citenamefont {Bao},
  \citenamefont {Duan}, \citenamefont {Jin}, \citenamefont {Lu}, \citenamefont
  {Li}, \citenamefont {Qu}, \citenamefont {Wang}, \citenamefont {Novikova},
  \citenamefont {Mikhailov}, \citenamefont {Zhao}, \citenamefont {M{\o}lmer},
  \citenamefont {Shen},\ and\ \citenamefont {Xiao}}]{bao2020Nature}%
  \BibitemOpen
  \bibfield  {author} {\bibinfo {author} {\bibfnamefont {H.}~\bibnamefont
  {Bao}}, \bibinfo {author} {\bibfnamefont {J.~L.}\ \bibnamefont {Duan}},
  \bibinfo {author} {\bibfnamefont {S.~C.}\ \bibnamefont {Jin}}, \bibinfo
  {author} {\bibfnamefont {X.~D.}\ \bibnamefont {Lu}}, \bibinfo {author}
  {\bibfnamefont {P.~X.}\ \bibnamefont {Li}}, \bibinfo {author} {\bibfnamefont
  {W.~Z.}\ \bibnamefont {Qu}}, \bibinfo {author} {\bibfnamefont {M.~F.}\
  \bibnamefont {Wang}}, \bibinfo {author} {\bibfnamefont {I.}~\bibnamefont
  {Novikova}}, \bibinfo {author} {\bibfnamefont {E.~E.}\ \bibnamefont
  {Mikhailov}}, \bibinfo {author} {\bibfnamefont {K.~F.}\ \bibnamefont {Zhao}},
  \bibinfo {author} {\bibfnamefont {K.}~\bibnamefont {M{\o}lmer}}, \bibinfo
  {author} {\bibfnamefont {H.}~\bibnamefont {Shen}}, \ and\ \bibinfo {author}
  {\bibfnamefont {Y.~H.}\ \bibnamefont {Xiao}},\ }\bibfield  {title} {\enquote
  {\bibinfo {title} {Spin squeezing of $10^{11}$ atoms by prediction and
  retrodiction measurements},}\ }\href {\doibase
  https://doi.org/10.1038/s41586-020-2243-7} {\bibfield  {journal} {\bibinfo
  {journal} {Nature}\ }\textbf {\bibinfo {volume} {581}},\ \bibinfo {pages}
  {159} (\bibinfo {year} {2020})}\BibitemShut {NoStop}%
\bibitem [{\citenamefont {Hardman}\ \emph {et~al.}(2016)\citenamefont
  {Hardman}, \citenamefont {Everitt}, \citenamefont {McDonald}, \citenamefont
  {Manju}, \citenamefont {Wigley}, \citenamefont {Sooriyabandara},
  \citenamefont {Kuhn}, \citenamefont {Debs}, \citenamefont {Close},\ and\
  \citenamefont {Robins}}]{hardman2016PRL}%
  \BibitemOpen
  \bibfield  {author} {\bibinfo {author} {\bibfnamefont {K.~S.}\ \bibnamefont
  {Hardman}}, \bibinfo {author} {\bibfnamefont {P.~J.}\ \bibnamefont
  {Everitt}}, \bibinfo {author} {\bibfnamefont {G.~D.}\ \bibnamefont
  {McDonald}}, \bibinfo {author} {\bibfnamefont {P.}~\bibnamefont {Manju}},
  \bibinfo {author} {\bibfnamefont {P.~B.}\ \bibnamefont {Wigley}}, \bibinfo
  {author} {\bibfnamefont {M.~A.}\ \bibnamefont {Sooriyabandara}}, \bibinfo
  {author} {\bibfnamefont {C.~C.~N.}\ \bibnamefont {Kuhn}}, \bibinfo {author}
  {\bibfnamefont {J.~E.}\ \bibnamefont {Debs}}, \bibinfo {author}
  {\bibfnamefont {J.~D.}\ \bibnamefont {Close}}, \ and\ \bibinfo {author}
  {\bibfnamefont {N.~P.}\ \bibnamefont {Robins}},\ }\bibfield  {title}
  {\enquote {\bibinfo {title} {Simultaneous precision gravimetry and magnetic
  gradiometry with a {Bose-Einstein} condensate: A high precision, quantum
  sensor},}\ }\href {\doibase https://doi.org/10.1103/PhysRevLett.117.138501}
  {\bibfield  {journal} {\bibinfo  {journal} {Phys. Rev. Lett.}\ }\textbf
  {\bibinfo {volume} {117}},\ \bibinfo {pages} {138501} (\bibinfo {year}
  {2016})}\BibitemShut {NoStop}%
\bibitem [{\citenamefont {Asenbaum}\ \emph {et~al.}(2017)\citenamefont
  {Asenbaum}, \citenamefont {Overstreet}, \citenamefont {Kovachy},
  \citenamefont {Brown}, \citenamefont {Hogan},\ and\ \citenamefont
  {Kasevich}}]{asenbaum2017PRL}%
  \BibitemOpen
  \bibfield  {author} {\bibinfo {author} {\bibfnamefont {P.}~\bibnamefont
  {Asenbaum}}, \bibinfo {author} {\bibfnamefont {C.}~\bibnamefont
  {Overstreet}}, \bibinfo {author} {\bibfnamefont {T.}~\bibnamefont {Kovachy}},
  \bibinfo {author} {\bibfnamefont {D.~D.}\ \bibnamefont {Brown}}, \bibinfo
  {author} {\bibfnamefont {J.~M.}\ \bibnamefont {Hogan}}, \ and\ \bibinfo
  {author} {\bibfnamefont {M.~A.}\ \bibnamefont {Kasevich}},\ }\bibfield
  {title} {\enquote {\bibinfo {title} {Phase shift in an atom interferometer
  due to spacetime curvature across its wave function},}\ }\href {\doibase
  https://doi.org/10.1103/PhysRevLett.118.183602} {\bibfield  {journal}
  {\bibinfo  {journal} {Phys. Rev. Lett.}\ }\textbf {\bibinfo {volume} {118}},\
  \bibinfo {pages} {183602} (\bibinfo {year} {2017})}\BibitemShut {NoStop}%
\bibitem [{\citenamefont {Cai}\ and\ \citenamefont
  {Plenio}(2013)}]{cai2013PRL}%
  \BibitemOpen
  \bibfield  {author} {\bibinfo {author} {\bibfnamefont {J.~M.}\ \bibnamefont
  {Cai}}\ and\ \bibinfo {author} {\bibfnamefont {M.~B.}\ \bibnamefont
  {Plenio}},\ }\bibfield  {title} {\enquote {\bibinfo {title} {Chemical compass
  model for avian magnetoreception as a quantum coherent device},}\ }\href
  {\doibase https://doi.org/10.1103/PhysRevLett.111.230503} {\bibfield
  {journal} {\bibinfo  {journal} {Phys. Rev. Lett.}\ }\textbf {\bibinfo
  {volume} {111}},\ \bibinfo {pages} {230503} (\bibinfo {year}
  {2013})}\BibitemShut {NoStop}%
\bibitem [{\citenamefont {Taylor}\ and\ \citenamefont
  {Bowen}(2016)}]{taylor2016PR}%
  \BibitemOpen
  \bibfield  {author} {\bibinfo {author} {\bibfnamefont {M.~A.}\ \bibnamefont
  {Taylor}}\ and\ \bibinfo {author} {\bibfnamefont {W.~P.}\ \bibnamefont
  {Bowen}},\ }\bibfield  {title} {\enquote {\bibinfo {title} {Quantum metrology
  and its application in biology},}\ }\href {\doibase
  https://doi.org/10.1016/j.physrep.2015.12.002} {\bibfield  {journal}
  {\bibinfo  {journal} {Phys. Rep.}\ }\textbf {\bibinfo {volume} {615}},\
  \bibinfo {pages} {1} (\bibinfo {year} {2016})}\BibitemShut {NoStop}%
\bibitem [{\citenamefont {Crespi}\ \emph {et~al.}(2012)\citenamefont {Crespi},
  \citenamefont {Lobino}, \citenamefont {Matthews}, \citenamefont {Politi},
  \citenamefont {Neal}, \citenamefont {Ramponi}, \citenamefont {Osellame},\
  and\ \citenamefont {O'Brien}}]{crespi2012APL}%
  \BibitemOpen
  \bibfield  {author} {\bibinfo {author} {\bibfnamefont {A.}~\bibnamefont
  {Crespi}}, \bibinfo {author} {\bibfnamefont {M.}~\bibnamefont {Lobino}},
  \bibinfo {author} {\bibfnamefont {J.~C.~F.}\ \bibnamefont {Matthews}},
  \bibinfo {author} {\bibfnamefont {A.}~\bibnamefont {Politi}}, \bibinfo
  {author} {\bibfnamefont {C.~R.}\ \bibnamefont {Neal}}, \bibinfo {author}
  {\bibfnamefont {R.}~\bibnamefont {Ramponi}}, \bibinfo {author} {\bibfnamefont
  {R.}~\bibnamefont {Osellame}}, \ and\ \bibinfo {author} {\bibfnamefont
  {J.~L.}\ \bibnamefont {O'Brien}},\ }\bibfield  {title} {\enquote {\bibinfo
  {title} {Measuring protein concentration with entangled photons},}\ }\href
  {\doibase https://doi.org/10.1063/1.4724105} {\bibfield  {journal} {\bibinfo
  {journal} {Appl. Phys. Lett}\ }\textbf {\bibinfo {volume} {100}},\ \bibinfo
  {pages} {233704} (\bibinfo {year} {2012})}\BibitemShut {NoStop}%
\bibitem [{\citenamefont {Taylor}\ \emph {et~al.}(2014)\citenamefont {Taylor},
  \citenamefont {Janousek}, \citenamefont {Daria}, \citenamefont {Knittel},
  \citenamefont {Hage}, \citenamefont {Bachor},\ and\ \citenamefont
  {Bowen}}]{taylor2014PRX}%
  \BibitemOpen
  \bibfield  {author} {\bibinfo {author} {\bibfnamefont {M.~A.}\ \bibnamefont
  {Taylor}}, \bibinfo {author} {\bibfnamefont {J.}~\bibnamefont {Janousek}},
  \bibinfo {author} {\bibfnamefont {V.}~\bibnamefont {Daria}}, \bibinfo
  {author} {\bibfnamefont {J.}~\bibnamefont {Knittel}}, \bibinfo {author}
  {\bibfnamefont {B.}~\bibnamefont {Hage}}, \bibinfo {author} {\bibfnamefont
  {H.-A.}\ \bibnamefont {Bachor}}, \ and\ \bibinfo {author} {\bibfnamefont
  {W.~P.}\ \bibnamefont {Bowen}},\ }\bibfield  {title} {\enquote {\bibinfo
  {title} {Subdiffraction-limited quantum imaging within a living cell},}\
  }\href {\doibase https://doi.org/10.1103/PhysRevX.4.011017} {\bibfield
  {journal} {\bibinfo  {journal} {Phys. Rev. X}\ }\textbf {\bibinfo {volume}
  {4}},\ \bibinfo {pages} {011017} (\bibinfo {year} {2014})}\BibitemShut
  {NoStop}%
\bibitem [{\citenamefont {Cai}\ \emph {et~al.}(2012)\citenamefont {Cai},
  \citenamefont {Ai}, \citenamefont {Quan},\ and\ \citenamefont
  {Sun}}]{cai2012PRA}%
  \BibitemOpen
  \bibfield  {author} {\bibinfo {author} {\bibfnamefont {C.~Y.}\ \bibnamefont
  {Cai}}, \bibinfo {author} {\bibfnamefont {Q.}~\bibnamefont {Ai}}, \bibinfo
  {author} {\bibfnamefont {H.~T.}\ \bibnamefont {Quan}}, \ and\ \bibinfo
  {author} {\bibfnamefont {C.~P.}\ \bibnamefont {Sun}},\ }\bibfield  {title}
  {\enquote {\bibinfo {title} {Sensitive chemical compass assisted by quantum
  criticality},}\ }\href {\doibase https://doi.org/10.1103/PhysRevA.85.022315}
  {\bibfield  {journal} {\bibinfo  {journal} {Phys. Rev. A}\ }\textbf {\bibinfo
  {volume} {85}},\ \bibinfo {pages} {022315} (\bibinfo {year}
  {2012})}\BibitemShut {NoStop}%
\bibitem [{\citenamefont {Yang}\ \emph {et~al.}(2012)\citenamefont {Yang},
  \citenamefont {Ai},\ and\ \citenamefont {Sun}}]{yang2012PRA}%
  \BibitemOpen
  \bibfield  {author} {\bibinfo {author} {\bibfnamefont {L.~P.}\ \bibnamefont
  {Yang}}, \bibinfo {author} {\bibfnamefont {Q.}~\bibnamefont {Ai}}, \ and\
  \bibinfo {author} {\bibfnamefont {C.~P.}\ \bibnamefont {Sun}},\ }\bibfield
  {title} {\enquote {\bibinfo {title} {Generalized {H}olstein model for
  spin-dependent electron-transfer reactions},}\ }\href {\doibase
  https://doi.org/10.1103/PhysRevA.85.032707} {\bibfield  {journal} {\bibinfo
  {journal} {Phys. Rev. A}\ }\textbf {\bibinfo {volume} {85}},\ \bibinfo
  {pages} {032707} (\bibinfo {year} {2012})}\BibitemShut {NoStop}%
\bibitem [{\citenamefont {Giovannetti}\ \emph {et~al.}(2011)\citenamefont
  {Giovannetti}, \citenamefont {Lloyd},\ and\ \citenamefont
  {Maccone}}]{Giovannetti2011NP}%
  \BibitemOpen
  \bibfield  {author} {\bibinfo {author} {\bibfnamefont {V.}~\bibnamefont
  {Giovannetti}}, \bibinfo {author} {\bibfnamefont {S.}~\bibnamefont {Lloyd}},
  \ and\ \bibinfo {author} {\bibfnamefont {L.}~\bibnamefont {Maccone}},\
  }\bibfield  {title} {\enquote {\bibinfo {title} {Advances in quantum
  metrology},}\ }\href {\doibase https://doi.org/10.1038/nphoton.2011.35}
  {\bibfield  {journal} {\bibinfo  {journal} {Nat. Photon.}\ }\textbf {\bibinfo
  {volume} {5}},\ \bibinfo {pages} {222} (\bibinfo {year} {2011})}\BibitemShut
  {NoStop}%
\bibitem [{\citenamefont {Wang}\ \emph {et~al.}(2017)\citenamefont {Wang},
  \citenamefont {Chen},\ and\ \citenamefont {An}}]{Wang2017NJP}%
  \BibitemOpen
  \bibfield  {author} {\bibinfo {author} {\bibfnamefont {Y.~S.}\ \bibnamefont
  {Wang}}, \bibinfo {author} {\bibfnamefont {C.}~\bibnamefont {Chen}}, \ and\
  \bibinfo {author} {\bibfnamefont {J.~H.}\ \bibnamefont {An}},\ }\bibfield
  {title} {\enquote {\bibinfo {title} {Quantum metrology in local dissipative
  environments},}\ }\href {\doibase 10.1088/1367-2630/aa8b01} {\bibfield
  {journal} {\bibinfo  {journal} {New J. Phys.}\ }\textbf {\bibinfo {volume}
  {19}},\ \bibinfo {pages} {113019} (\bibinfo {year} {2017})}\BibitemShut
  {NoStop}%
\bibitem [{\citenamefont {Szigeti}\ \emph {et~al.}(2020)\citenamefont
  {Szigeti}, \citenamefont {Nolan}, \citenamefont {Close},\ and\ \citenamefont
  {Haine}}]{szigeti2020PRL}%
  \BibitemOpen
  \bibfield  {author} {\bibinfo {author} {\bibfnamefont {S.~S.}\ \bibnamefont
  {Szigeti}}, \bibinfo {author} {\bibfnamefont {S.~P.}\ \bibnamefont {Nolan}},
  \bibinfo {author} {\bibfnamefont {J.~D.}\ \bibnamefont {Close}}, \ and\
  \bibinfo {author} {\bibfnamefont {S.~A.}\ \bibnamefont {Haine}},\ }\bibfield
  {title} {\enquote {\bibinfo {title} {High-precision quantum-enhanced
  gravimetry with a {Bose-Einstein} condensate},}\ }\href {\doibase
  https://doi.org/10.1103/PhysRevLett.125.100402} {\bibfield  {journal}
  {\bibinfo  {journal} {Phys. Rev. Lett.}\ }\textbf {\bibinfo {volume} {125}},\
  \bibinfo {pages} {100402} (\bibinfo {year} {2020})}\BibitemShut {NoStop}%
\bibitem [{\citenamefont {Luo}\ \emph {et~al.}(2017)\citenamefont {Luo},
  \citenamefont {Huang}, \citenamefont {Zhang},\ and\ \citenamefont
  {Lee}}]{luo2017PRA}%
  \BibitemOpen
  \bibfield  {author} {\bibinfo {author} {\bibfnamefont {C.~Y.}\ \bibnamefont
  {Luo}}, \bibinfo {author} {\bibfnamefont {J.~H.}\ \bibnamefont {Huang}},
  \bibinfo {author} {\bibfnamefont {X.~D.}\ \bibnamefont {Zhang}}, \ and\
  \bibinfo {author} {\bibfnamefont {C.~H.}\ \bibnamefont {Lee}},\ }\bibfield
  {title} {\enquote {\bibinfo {title} {Heisenberg-limited {Sagnac}
  interferometer with multiparticle states},}\ }\href {\doibase
  https://doi.org/10.1103/PhysRevA.95.023608} {\bibfield  {journal} {\bibinfo
  {journal} {Phys. Rev. A}\ }\textbf {\bibinfo {volume} {95}},\ \bibinfo
  {pages} {023608} (\bibinfo {year} {2017})}\BibitemShut {NoStop}%
\bibitem [{\citenamefont {Pezz{\'e}}\ and\ \citenamefont
  {Smerzi}(2009)}]{pezze2009PRL}%
  \BibitemOpen
  \bibfield  {author} {\bibinfo {author} {\bibfnamefont {L.}~\bibnamefont
  {Pezz{\'e}}}\ and\ \bibinfo {author} {\bibfnamefont {A.}~\bibnamefont
  {Smerzi}},\ }\bibfield  {title} {\enquote {\bibinfo {title} {Entanglement,
  nonlinear dynamics, and the {Heisenberg} limit},}\ }\href {\doibase
  https://doi.org/10.1103/PhysRevLett.102.100401} {\bibfield  {journal}
  {\bibinfo  {journal} {Phys. Rev. Lett.}\ }\textbf {\bibinfo {volume} {102}},\
  \bibinfo {pages} {100401} (\bibinfo {year} {2009})}\BibitemShut {NoStop}%
\bibitem [{\citenamefont {Joo}\ \emph {et~al.}(2011)\citenamefont {Joo},
  \citenamefont {Munro},\ and\ \citenamefont {Spiller}}]{joo2011PRL}%
  \BibitemOpen
  \bibfield  {author} {\bibinfo {author} {\bibfnamefont {J.}~\bibnamefont
  {Joo}}, \bibinfo {author} {\bibfnamefont {W.~J.}\ \bibnamefont {Munro}}, \
  and\ \bibinfo {author} {\bibfnamefont {T.~P.}\ \bibnamefont {Spiller}},\
  }\bibfield  {title} {\enquote {\bibinfo {title} {Quantum metrology with
  entangled coherent states},}\ }\href {\doibase
  https://doi.org/10.1103/PhysRevLett.107.083601} {\bibfield  {journal}
  {\bibinfo  {journal} {Phys. Rev. Lett.}\ }\textbf {\bibinfo {volume} {107}},\
  \bibinfo {pages} {083601} (\bibinfo {year} {2011})}\BibitemShut {NoStop}%
\bibitem [{\citenamefont {He}\ \emph {et~al.}(2021)\citenamefont {He},
  \citenamefont {Guang}, \citenamefont {Li}, \citenamefont {Deng},
  \citenamefont {Zhang}, \citenamefont {Zhao}, \citenamefont {Deng},\ and\
  \citenamefont {Ai}}]{he2021PRA}%
  \BibitemOpen
  \bibfield  {author} {\bibinfo {author} {\bibfnamefont {W.~T.}\ \bibnamefont
  {He}}, \bibinfo {author} {\bibfnamefont {H.~Y.}\ \bibnamefont {Guang}},
  \bibinfo {author} {\bibfnamefont {Z.~Y.}\ \bibnamefont {Li}}, \bibinfo
  {author} {\bibfnamefont {R.~Q.}\ \bibnamefont {Deng}}, \bibinfo {author}
  {\bibfnamefont {N.~N.}\ \bibnamefont {Zhang}}, \bibinfo {author}
  {\bibfnamefont {J.~X.}\ \bibnamefont {Zhao}}, \bibinfo {author}
  {\bibfnamefont {F.~G.}\ \bibnamefont {Deng}}, \ and\ \bibinfo {author}
  {\bibfnamefont {Q.}~\bibnamefont {Ai}},\ }\bibfield  {title} {\enquote
  {\bibinfo {title} {Quantum metrology with one auxiliary particle in a
  correlated bath and its quantum simulation},}\ }\href {\doibase
  https://doi.org/10.1103/PhysRevA.104.062429} {\bibfield  {journal} {\bibinfo
  {journal} {Phys. Rev. A}\ }\textbf {\bibinfo {volume} {104}},\ \bibinfo
  {pages} {062429} (\bibinfo {year} {2021})}\BibitemShut {NoStop}%
\bibitem [{\citenamefont {Zhou}\ \emph {et~al.}(2018)\citenamefont {Zhou},
  \citenamefont {Zhang}, \citenamefont {Preskill},\ and\ \citenamefont
  {Jiang}}]{zhou2018NC}%
  \BibitemOpen
  \bibfield  {author} {\bibinfo {author} {\bibfnamefont {S.~S.}\ \bibnamefont
  {Zhou}}, \bibinfo {author} {\bibfnamefont {M.~Z.}\ \bibnamefont {Zhang}},
  \bibinfo {author} {\bibfnamefont {J.}~\bibnamefont {Preskill}}, \ and\
  \bibinfo {author} {\bibfnamefont {L.}~\bibnamefont {Jiang}},\ }\bibfield
  {title} {\enquote {\bibinfo {title} {Achieving the {Heisenberg} limit in
  quantum metrology using quantum error correction},}\ }\href {\doibase
  https://doi.org/10.1038/s41467-017-02510-3} {\bibfield  {journal} {\bibinfo
  {journal} {Nat. Commun.}\ }\textbf {\bibinfo {volume} {9}},\ \bibinfo {pages}
  {78} (\bibinfo {year} {2018})}\BibitemShut {NoStop}%
\bibitem [{\citenamefont {Ma}\ \emph {et~al.}(2011)\citenamefont {Ma},
  \citenamefont {Wang}, \citenamefont {Sun},\ and\ \citenamefont
  {Nori}}]{ma2011PR}%
  \BibitemOpen
  \bibfield  {author} {\bibinfo {author} {\bibfnamefont {J.}~\bibnamefont
  {Ma}}, \bibinfo {author} {\bibfnamefont {X.~G.}\ \bibnamefont {Wang}},
  \bibinfo {author} {\bibfnamefont {C.~P.}\ \bibnamefont {Sun}}, \ and\
  \bibinfo {author} {\bibfnamefont {F.}~\bibnamefont {Nori}},\ }\bibfield
  {title} {\enquote {\bibinfo {title} {Quantum spin squeezing},}\ }\href
  {\doibase https://doi.org/10.1016/j.physrep.2011.08.003} {\bibfield
  {journal} {\bibinfo  {journal} {Phys. Rep.}\ }\textbf {\bibinfo {volume}
  {509}},\ \bibinfo {pages} {89} (\bibinfo {year} {2011})}\BibitemShut
  {NoStop}%
\bibitem [{\citenamefont {Engelsen}\ \emph {et~al.}(2017)\citenamefont
  {Engelsen}, \citenamefont {Krishnakumar}, \citenamefont {Hosten},\ and\
  \citenamefont {Kasevich}}]{engelsen2017PRL}%
  \BibitemOpen
  \bibfield  {author} {\bibinfo {author} {\bibfnamefont {N.~J.}\ \bibnamefont
  {Engelsen}}, \bibinfo {author} {\bibfnamefont {R.}~\bibnamefont
  {Krishnakumar}}, \bibinfo {author} {\bibfnamefont {O.}~\bibnamefont
  {Hosten}}, \ and\ \bibinfo {author} {\bibfnamefont {M.~A.}\ \bibnamefont
  {Kasevich}},\ }\bibfield  {title} {\enquote {\bibinfo {title} {Bell
  correlations in spin-squeezed states of 500 000 atoms},}\ }\href {\doibase
  https://doi.org/10.1103/PhysRevLett.118.140401} {\bibfield  {journal}
  {\bibinfo  {journal} {Phys. Rev. Lett.}\ }\textbf {\bibinfo {volume} {118}},\
  \bibinfo {pages} {140401} (\bibinfo {year} {2017})}\BibitemShut {NoStop}%
\bibitem [{\citenamefont {Holland}\ and\ \citenamefont
  {Burnett}(1993)}]{holland1993PRL}%
  \BibitemOpen
  \bibfield  {author} {\bibinfo {author} {\bibfnamefont {M.~J.}\ \bibnamefont
  {Holland}}\ and\ \bibinfo {author} {\bibfnamefont {K.}~\bibnamefont
  {Burnett}},\ }\bibfield  {title} {\enquote {\bibinfo {title} {Interferometric
  detection of optical phase shifts at the {H}eisenberg limit},}\ }\href
  {\doibase https://doi.org/10.1103/PhysRevLett.71.1355} {\bibfield  {journal}
  {\bibinfo  {journal} {Phys. Rev. Lett.}\ }\textbf {\bibinfo {volume} {71}},\
  \bibinfo {pages} {1355} (\bibinfo {year} {1993})}\BibitemShut {NoStop}%
\bibitem [{\citenamefont {Huelga}\ \emph {et~al.}(1997)\citenamefont {Huelga},
  \citenamefont {Macchiavello}, \citenamefont {Pellizzari},\ and\ \citenamefont
  {Ekert}}]{Huelga1997PRL}%
  \BibitemOpen
  \bibfield  {author} {\bibinfo {author} {\bibfnamefont {S.~F.}\ \bibnamefont
  {Huelga}}, \bibinfo {author} {\bibfnamefont {C.}~\bibnamefont
  {Macchiavello}}, \bibinfo {author} {\bibfnamefont {T.}~\bibnamefont
  {Pellizzari}}, \ and\ \bibinfo {author} {\bibfnamefont {A.~K.}\ \bibnamefont
  {Ekert}},\ }\bibfield  {title} {\enquote {\bibinfo {title} {Improvement of
  frequency standards with quantum entanglement},}\ }\href {\doibase
  https://doi.org/10.1103/PhysRevLett.79.3865} {\bibfield  {journal} {\bibinfo
  {journal} {Phys. Rev. Lett.}\ }\textbf {\bibinfo {volume} {79}},\ \bibinfo
  {pages} {3865} (\bibinfo {year} {1997})}\BibitemShut {NoStop}%
\bibitem [{\citenamefont {Chin}\ \emph {et~al.}(2012)\citenamefont {Chin},
  \citenamefont {Huelga},\ and\ \citenamefont {Plenio}}]{chin2012PRL}%
  \BibitemOpen
  \bibfield  {author} {\bibinfo {author} {\bibfnamefont {A.~W.}\ \bibnamefont
  {Chin}}, \bibinfo {author} {\bibfnamefont {S.~F.}\ \bibnamefont {Huelga}}, \
  and\ \bibinfo {author} {\bibfnamefont {M.~B.}\ \bibnamefont {Plenio}},\
  }\bibfield  {title} {\enquote {\bibinfo {title} {Quantum metrology in
  {non-Markovian} environments},}\ }\href {\doibase
  https://doi.org/10.1103/PhysRevLett.109.233601} {\bibfield  {journal}
  {\bibinfo  {journal} {Phys. Rev. Lett.}\ }\textbf {\bibinfo {volume} {109}},\
  \bibinfo {pages} {233601} (\bibinfo {year} {2012})}\BibitemShut {NoStop}%
\bibitem [{\citenamefont {Matsuzaki}\ \emph {et~al.}(2011)\citenamefont
  {Matsuzaki}, \citenamefont {Benjamin},\ and\ \citenamefont
  {Fitzsimons}}]{Matsuzaki2011PRA}%
  \BibitemOpen
  \bibfield  {author} {\bibinfo {author} {\bibfnamefont {Y.}~\bibnamefont
  {Matsuzaki}}, \bibinfo {author} {\bibfnamefont {S.~C.}\ \bibnamefont
  {Benjamin}}, \ and\ \bibinfo {author} {\bibfnamefont {J.}~\bibnamefont
  {Fitzsimons}},\ }\bibfield  {title} {\enquote {\bibinfo {title} {Magnetic
  field sensing beyond the standard quantum limit under the effect of
  decoherence},}\ }\href {\doibase https://doi.org/10.1103/PhysRevA.84.012103}
  {\bibfield  {journal} {\bibinfo  {journal} {Phys. Rev. A}\ }\textbf {\bibinfo
  {volume} {84}},\ \bibinfo {pages} {012103} (\bibinfo {year}
  {2011})}\BibitemShut {NoStop}%
\bibitem [{\citenamefont {Long}\ \emph {et~al.}(2022)\citenamefont {Long},
  \citenamefont {He}, \citenamefont {Zhang}, \citenamefont {Tang},
  \citenamefont {Lin}, \citenamefont {Liu}, \citenamefont {Nie}, \citenamefont
  {Feng}, \citenamefont {Li}, \citenamefont {Xin}, \citenamefont {Ai},\ and\
  \citenamefont {Lu}}]{long2022PRL}%
  \BibitemOpen
  \bibfield  {author} {\bibinfo {author} {\bibfnamefont {X.~Y.}\ \bibnamefont
  {Long}}, \bibinfo {author} {\bibfnamefont {W.~T.}\ \bibnamefont {He}},
  \bibinfo {author} {\bibfnamefont {N.~N.}\ \bibnamefont {Zhang}}, \bibinfo
  {author} {\bibfnamefont {K.}~\bibnamefont {Tang}}, \bibinfo {author}
  {\bibfnamefont {Z.~D.}\ \bibnamefont {Lin}}, \bibinfo {author} {\bibfnamefont
  {H.~F.}\ \bibnamefont {Liu}}, \bibinfo {author} {\bibfnamefont {X.~F.}\
  \bibnamefont {Nie}}, \bibinfo {author} {\bibfnamefont {G.~R.}\ \bibnamefont
  {Feng}}, \bibinfo {author} {\bibfnamefont {J.}~\bibnamefont {Li}}, \bibinfo
  {author} {\bibfnamefont {T.}~\bibnamefont {Xin}}, \bibinfo {author}
  {\bibfnamefont {Q.}~\bibnamefont {Ai}}, \ and\ \bibinfo {author}
  {\bibfnamefont {D.-W.}\ \bibnamefont {Lu}},\ }\bibfield  {title} {\enquote
  {\bibinfo {title} {Entanglement-enhanced quantum metrology in colored noise
  by quantum {Zeno} effect},}\ }\href {\doibase
  https://doi.org/10.1103/PhysRevLett.129.070502} {\bibfield  {journal}
  {\bibinfo  {journal} {Phys. Rev. Lett.}\ }\textbf {\bibinfo {volume} {129}},\
  \bibinfo {pages} {070502} (\bibinfo {year} {2022})}\BibitemShut {NoStop}%
\bibitem [{\citenamefont {Bai}\ and\ \citenamefont {An}(2023)}]{bai2023PRL}%
  \BibitemOpen
  \bibfield  {author} {\bibinfo {author} {\bibfnamefont {S.~Y.}\ \bibnamefont
  {Bai}}\ and\ \bibinfo {author} {\bibfnamefont {J.~H.}\ \bibnamefont {An}},\
  }\bibfield  {title} {\enquote {\bibinfo {title} {Floquet engineering to
  overcome no-go theorem of noisy quantum metrology},}\ }\href {\doibase
  https://doi.org/10.1103/PhysRevLett.131.050801} {\bibfield  {journal}
  {\bibinfo  {journal} {Phys. Rev. Lett.}\ }\textbf {\bibinfo {volume} {131}},\
  \bibinfo {pages} {050801} (\bibinfo {year} {2023})}\BibitemShut {NoStop}%
\bibitem [{\citenamefont {Smirne}\ \emph {et~al.}(2016)\citenamefont {Smirne},
  \citenamefont {Ko{\l}ody{\'n}ski}, \citenamefont {Huelga},\ and\
  \citenamefont {Demkowicz-Dobrza{\'n}ski}}]{smirne2016PRL}%
  \BibitemOpen
  \bibfield  {author} {\bibinfo {author} {\bibfnamefont {A.}~\bibnamefont
  {Smirne}}, \bibinfo {author} {\bibfnamefont {J.}~\bibnamefont
  {Ko{\l}ody{\'n}ski}}, \bibinfo {author} {\bibfnamefont {S.~F.}\ \bibnamefont
  {Huelga}}, \ and\ \bibinfo {author} {\bibfnamefont {R.}~\bibnamefont
  {Demkowicz-Dobrza{\'n}ski}},\ }\bibfield  {title} {\enquote {\bibinfo {title}
  {Ultimate precision limits for noisy frequency estimation},}\ }\href
  {\doibase https://doi.org/10.1103/PhysRevLett.116.120801} {\bibfield
  {journal} {\bibinfo  {journal} {Phys. Rev. Lett.}\ }\textbf {\bibinfo
  {volume} {116}},\ \bibinfo {pages} {120801} (\bibinfo {year}
  {2016})}\BibitemShut {NoStop}%
\bibitem [{\citenamefont {Macieszczak}(2015)}]{Macieszczak2015PRA}%
  \BibitemOpen
  \bibfield  {author} {\bibinfo {author} {\bibfnamefont {K.}~\bibnamefont
  {Macieszczak}},\ }\bibfield  {title} {\enquote {\bibinfo {title} {Zeno limit
  in frequency estimation with {non-Markovian} environments},}\ }\href
  {\doibase https://doi.org/10.1103/PhysRevA.92.010102} {\bibfield  {journal}
  {\bibinfo  {journal} {Phys. Rev. A}\ }\textbf {\bibinfo {volume} {92}},\
  \bibinfo {pages} {010102} (\bibinfo {year} {2015})}\BibitemShut {NoStop}%
\bibitem [{\citenamefont {Yamamoto}\ \emph {et~al.}(2022)\citenamefont
  {Yamamoto}, \citenamefont {Endo}, \citenamefont {Hakoshima}, \citenamefont
  {Matsuzaki},\ and\ \citenamefont {Tokunaga}}]{Yamamoto2022PRL}%
  \BibitemOpen
  \bibfield  {author} {\bibinfo {author} {\bibfnamefont {K.}~\bibnamefont
  {Yamamoto}}, \bibinfo {author} {\bibfnamefont {S.}~\bibnamefont {Endo}},
  \bibinfo {author} {\bibfnamefont {H.}~\bibnamefont {Hakoshima}}, \bibinfo
  {author} {\bibfnamefont {Y.}~\bibnamefont {Matsuzaki}}, \ and\ \bibinfo
  {author} {\bibfnamefont {Y.}~\bibnamefont {Tokunaga}},\ }\bibfield  {title}
  {\enquote {\bibinfo {title} {Error-mitigated quantum metrology via virtual
  purification},}\ }\href {\doibase 10.1103/PhysRevLett.129.250503} {\bibfield
  {journal} {\bibinfo  {journal} {Phys. Rev. Lett.}\ }\textbf {\bibinfo
  {volume} {129}},\ \bibinfo {pages} {250503} (\bibinfo {year}
  {2022})}\BibitemShut {NoStop}%
\bibitem [{\citenamefont {D{\"ur}}\ \emph {et~al.}(2014)\citenamefont
  {D{\"ur}}, \citenamefont {Skotiniotis}, \citenamefont {Fr\"owis},\ and\
  \citenamefont {Kraus}}]{Dur2014PRL}%
  \BibitemOpen
  \bibfield  {author} {\bibinfo {author} {\bibfnamefont {W.}~\bibnamefont
  {D{\"ur}}}, \bibinfo {author} {\bibfnamefont {M.}~\bibnamefont
  {Skotiniotis}}, \bibinfo {author} {\bibfnamefont {F.}~\bibnamefont
  {Fr\"owis}}, \ and\ \bibinfo {author} {\bibfnamefont {B.}~\bibnamefont
  {Kraus}},\ }\bibfield  {title} {\enquote {\bibinfo {title} {Improved quantum
  metrology using quantum error correction},}\ }\href {\doibase
  10.1103/PhysRevLett.112.080801} {\bibfield  {journal} {\bibinfo  {journal}
  {Phys. Rev. Lett.}\ }\textbf {\bibinfo {volume} {112}},\ \bibinfo {pages}
  {080801} (\bibinfo {year} {2014})}\BibitemShut {NoStop}%
\bibitem [{\citenamefont {Rojkov}\ \emph {et~al.}(2022)\citenamefont {Rojkov},
  \citenamefont {Layden}, \citenamefont {Cappellaro}, \citenamefont {Home},\
  and\ \citenamefont {Reiter}}]{Rojkov2022PRL}%
  \BibitemOpen
  \bibfield  {author} {\bibinfo {author} {\bibfnamefont {I.}~\bibnamefont
  {Rojkov}}, \bibinfo {author} {\bibfnamefont {D.}~\bibnamefont {Layden}},
  \bibinfo {author} {\bibfnamefont {P.}~\bibnamefont {Cappellaro}}, \bibinfo
  {author} {\bibfnamefont {J.}~\bibnamefont {Home}}, \ and\ \bibinfo {author}
  {\bibfnamefont {F.}~\bibnamefont {Reiter}},\ }\bibfield  {title} {\enquote
  {\bibinfo {title} {Bias in error-corrected quantum sensing},}\ }\href
  {\doibase 10.1103/PhysRevLett.128.140503} {\bibfield  {journal} {\bibinfo
  {journal} {Phys. Rev. Lett.}\ }\textbf {\bibinfo {volume} {128}},\ \bibinfo
  {pages} {140503} (\bibinfo {year} {2022})}\BibitemShut {NoStop}%
\bibitem [{\citenamefont {Reiter}\ \emph {et~al.}(2017)\citenamefont {Reiter},
  \citenamefont {S{\o}rensen}, \citenamefont {Zoller},\ and\ \citenamefont
  {Muschik}}]{reiter2017NC}%
  \BibitemOpen
  \bibfield  {author} {\bibinfo {author} {\bibfnamefont {F.}~\bibnamefont
  {Reiter}}, \bibinfo {author} {\bibfnamefont {A.~S.}\ \bibnamefont
  {S{\o}rensen}}, \bibinfo {author} {\bibfnamefont {P.}~\bibnamefont {Zoller}},
  \ and\ \bibinfo {author} {\bibfnamefont {C.~A.}\ \bibnamefont {Muschik}},\
  }\bibfield  {title} {\enquote {\bibinfo {title} {Dissipative quantum error
  correction and application to quantum sensing with trapped ions},}\ }\href
  {\doibase https://doi.org/10.1038/s41467-017-01895-5} {\bibfield  {journal}
  {\bibinfo  {journal} {Nat. Commun.}\ }\textbf {\bibinfo {volume} {8}},\
  \bibinfo {pages} {1822} (\bibinfo {year} {2017})}\BibitemShut {NoStop}%
\bibitem [{\citenamefont {Rossi}\ \emph {et~al.}(2020)\citenamefont {Rossi},
  \citenamefont {Albarelli}, \citenamefont {Tamascelli},\ and\ \citenamefont
  {Genoni}}]{Rossi2020PRL}%
  \BibitemOpen
  \bibfield  {author} {\bibinfo {author} {\bibfnamefont {M.~A.~C.}\
  \bibnamefont {Rossi}}, \bibinfo {author} {\bibfnamefont {F.}~\bibnamefont
  {Albarelli}}, \bibinfo {author} {\bibfnamefont {D.}~\bibnamefont
  {Tamascelli}}, \ and\ \bibinfo {author} {\bibfnamefont {M.~G.}\ \bibnamefont
  {Genoni}},\ }\bibfield  {title} {\enquote {\bibinfo {title} {Noisy quantum
  metrology enhanced by continuous nondemolition measurement},}\ }\href
  {\doibase 10.1103/PhysRevLett.125.200505} {\bibfield  {journal} {\bibinfo
  {journal} {Phys. Rev. Lett.}\ }\textbf {\bibinfo {volume} {125}},\ \bibinfo
  {pages} {200505} (\bibinfo {year} {2020})}\BibitemShut {NoStop}%
\bibitem [{\citenamefont {Bai}\ \emph {et~al.}(2019)\citenamefont {Bai},
  \citenamefont {Peng}, \citenamefont {Luo},\ and\ \citenamefont
  {An}}]{bai2019PRL}%
  \BibitemOpen
  \bibfield  {author} {\bibinfo {author} {\bibfnamefont {K.}~\bibnamefont
  {Bai}}, \bibinfo {author} {\bibfnamefont {Z.}~\bibnamefont {Peng}}, \bibinfo
  {author} {\bibfnamefont {H.~G.}\ \bibnamefont {Luo}}, \ and\ \bibinfo
  {author} {\bibfnamefont {J.~H.}\ \bibnamefont {An}},\ }\bibfield  {title}
  {\enquote {\bibinfo {title} {Retrieving ideal precision in noisy quantum
  optical metrology},}\ }\href {\doibase 10.1103/PhysRevLett.123.040402}
  {\bibfield  {journal} {\bibinfo  {journal} {Phys. Rev. Lett.}\ }\textbf
  {\bibinfo {volume} {123}},\ \bibinfo {pages} {040402} (\bibinfo {year}
  {2019})}\BibitemShut {NoStop}%
\bibitem [{\citenamefont {Shi}\ \emph {et~al.}(2016)\citenamefont {Shi},
  \citenamefont {Wu}, \citenamefont {Gonz\'alez-Tudela},\ and\ \citenamefont
  {Cirac}}]{Shi2016PRX}%
  \BibitemOpen
  \bibfield  {author} {\bibinfo {author} {\bibfnamefont {T.}~\bibnamefont
  {Shi}}, \bibinfo {author} {\bibfnamefont {Y.~H.}\ \bibnamefont {Wu}},
  \bibinfo {author} {\bibfnamefont {A.}~\bibnamefont {Gonz\'alez-Tudela}}, \
  and\ \bibinfo {author} {\bibfnamefont {J.~I.}\ \bibnamefont {Cirac}},\
  }\bibfield  {title} {\enquote {\bibinfo {title} {Bound states in boson
  impurity models},}\ }\href {\doibase 10.1103/PhysRevX.6.021027} {\bibfield
  {journal} {\bibinfo  {journal} {Phys. Rev. X}\ }\textbf {\bibinfo {volume}
  {6}},\ \bibinfo {pages} {021027} (\bibinfo {year} {2016})}\BibitemShut
  {NoStop}%
\bibitem [{\citenamefont {Bardeen}\ \emph {et~al.}(1957)\citenamefont
  {Bardeen}, \citenamefont {Cooper},\ and\ \citenamefont
  {Schrieffer}}]{Bardeen1957PR}%
  \BibitemOpen
  \bibfield  {author} {\bibinfo {author} {\bibfnamefont {J.}~\bibnamefont
  {Bardeen}}, \bibinfo {author} {\bibfnamefont {L.~N.}\ \bibnamefont {Cooper}},
  \ and\ \bibinfo {author} {\bibfnamefont {J.~R.}\ \bibnamefont {Schrieffer}},\
  }\bibfield  {title} {\enquote {\bibinfo {title} {Theory of
  superconductivity},}\ }\href {\doibase 10.1103/PhysRev.108.1175} {\bibfield
  {journal} {\bibinfo  {journal} {Phys. Rev.}\ }\textbf {\bibinfo {volume}
  {108}},\ \bibinfo {pages} {1175} (\bibinfo {year} {1957})}\BibitemShut
  {NoStop}%
\bibitem [{\citenamefont {Holstein}(1959)}]{Holstein1959AP}%
  \BibitemOpen
  \bibfield  {author} {\bibinfo {author} {\bibfnamefont {T.}~\bibnamefont
  {Holstein}},\ }\bibfield  {title} {\enquote {\bibinfo {title} {Studies of
  polaron motion: Part i. the molecular-crystal model},}\ }\href {\doibase
  https://doi.org/10.1016/0003-4916(59)90002-8} {\bibfield  {journal} {\bibinfo
   {journal} {Ann. Phys.}\ }\textbf {\bibinfo {volume} {8}},\ \bibinfo {pages}
  {325} (\bibinfo {year} {1959})}\BibitemShut {NoStop}%
\bibitem [{\citenamefont {Shen}\ and\ \citenamefont {Fan}(2005)}]{Shen2005PRL}%
  \BibitemOpen
  \bibfield  {author} {\bibinfo {author} {\bibfnamefont {J.~T.}\ \bibnamefont
  {Shen}}\ and\ \bibinfo {author} {\bibfnamefont {S.~H.}\ \bibnamefont {Fan}},\
  }\bibfield  {title} {\enquote {\bibinfo {title} {Coherent single photon
  transport in a one-dimensional waveguide coupled with superconducting quantum
  bits},}\ }\href {\doibase 10.1103/PhysRevLett.95.213001} {\bibfield
  {journal} {\bibinfo  {journal} {Phys. Rev. Lett.}\ }\textbf {\bibinfo
  {volume} {95}},\ \bibinfo {pages} {213001} (\bibinfo {year}
  {2005})}\BibitemShut {NoStop}%
\bibitem [{\citenamefont {Chang}\ \emph {et~al.}(2007)\citenamefont {Chang},
  \citenamefont {S{\o}rensen}, \citenamefont {Demler},\ and\ \citenamefont
  {Lukin}}]{chang2007NP}%
  \BibitemOpen
  \bibfield  {author} {\bibinfo {author} {\bibfnamefont {D.~E.}\ \bibnamefont
  {Chang}}, \bibinfo {author} {\bibfnamefont {A.~S.}\ \bibnamefont
  {S{\o}rensen}}, \bibinfo {author} {\bibfnamefont {E.~A.}\ \bibnamefont
  {Demler}}, \ and\ \bibinfo {author} {\bibfnamefont {M.~D.}\ \bibnamefont
  {Lukin}},\ }\bibfield  {title} {\enquote {\bibinfo {title} {A single-photon
  transistor using nanoscale surface plasmons},}\ }\href {\doibase
  https://doi.org/10.1038/nphys708} {\bibfield  {journal} {\bibinfo  {journal}
  {Nat. Phys.}\ }\textbf {\bibinfo {volume} {3}},\ \bibinfo {pages} {807}
  (\bibinfo {year} {2007})}\BibitemShut {NoStop}%
\bibitem [{\citenamefont {Zhou}\ \emph
  {et~al.}(2008{\natexlab{a}})\citenamefont {Zhou}, \citenamefont {Gong},
  \citenamefont {Liu}, \citenamefont {Sun},\ and\ \citenamefont
  {Nori}}]{zhou2008PRL}%
  \BibitemOpen
  \bibfield  {author} {\bibinfo {author} {\bibfnamefont {L.}~\bibnamefont
  {Zhou}}, \bibinfo {author} {\bibfnamefont {Z.~R.}\ \bibnamefont {Gong}},
  \bibinfo {author} {\bibfnamefont {Y.~X.}\ \bibnamefont {Liu}}, \bibinfo
  {author} {\bibfnamefont {C.~P.}\ \bibnamefont {Sun}}, \ and\ \bibinfo
  {author} {\bibfnamefont {F.}~\bibnamefont {Nori}},\ }\bibfield  {title}
  {\enquote {\bibinfo {title} {Controllable scattering of a single photon
  inside a one-dimensional resonator waveguide},}\ }\href {\doibase
  10.1103/PhysRevLett.101.100501} {\bibfield  {journal} {\bibinfo  {journal}
  {Phys. Rev. Lett.}\ }\textbf {\bibinfo {volume} {101}},\ \bibinfo {pages}
  {100501} (\bibinfo {year} {2008}{\natexlab{a}})}\BibitemShut {NoStop}%
\bibitem [{\citenamefont {Zhou}\ \emph {et~al.}(2013)\citenamefont {Zhou},
  \citenamefont {Yang}, \citenamefont {Li},\ and\ \citenamefont
  {Sun}}]{zhou2013PRL}%
  \BibitemOpen
  \bibfield  {author} {\bibinfo {author} {\bibfnamefont {L.}~\bibnamefont
  {Zhou}}, \bibinfo {author} {\bibfnamefont {L.~P.}\ \bibnamefont {Yang}},
  \bibinfo {author} {\bibfnamefont {Y.}~\bibnamefont {Li}}, \ and\ \bibinfo
  {author} {\bibfnamefont {C.~P.}\ \bibnamefont {Sun}},\ }\bibfield  {title}
  {\enquote {\bibinfo {title} {Quantum routing of single photons with a cyclic
  three-level system},}\ }\href {\doibase 10.1103/PhysRevLett.111.103604}
  {\bibfield  {journal} {\bibinfo  {journal} {Phys. Rev. Lett.}\ }\textbf
  {\bibinfo {volume} {111}},\ \bibinfo {pages} {103604} (\bibinfo {year}
  {2013})}\BibitemShut {NoStop}%
\bibitem [{\citenamefont {Zhou}\ \emph
  {et~al.}(2008{\natexlab{b}})\citenamefont {Zhou}, \citenamefont {Dong},
  \citenamefont {Liu}, \citenamefont {Sun},\ and\ \citenamefont
  {Nori}}]{zhou2008PRA}%
  \BibitemOpen
  \bibfield  {author} {\bibinfo {author} {\bibfnamefont {L.}~\bibnamefont
  {Zhou}}, \bibinfo {author} {\bibfnamefont {H.}~\bibnamefont {Dong}}, \bibinfo
  {author} {\bibfnamefont {Y.~X.}\ \bibnamefont {Liu}}, \bibinfo {author}
  {\bibfnamefont {C.~P.}\ \bibnamefont {Sun}}, \ and\ \bibinfo {author}
  {\bibfnamefont {F.}~\bibnamefont {Nori}},\ }\bibfield  {title} {\enquote
  {\bibinfo {title} {Quantum supercavity with atomic mirrors},}\ }\href
  {\doibase 10.1103/PhysRevA.78.063827} {\bibfield  {journal} {\bibinfo
  {journal} {Phys. Rev. A}\ }\textbf {\bibinfo {volume} {78}},\ \bibinfo
  {pages} {063827} (\bibinfo {year} {2008}{\natexlab{b}})}\BibitemShut
  {NoStop}%
\bibitem [{\citenamefont {Yao}\ and\ \citenamefont {Ai}(2023)}]{yao2023adp}%
  \BibitemOpen
  \bibfield  {author} {\bibinfo {author} {\bibfnamefont {Y.‐X.}\ \bibnamefont
  {Yao}}\ and\ \bibinfo {author} {\bibfnamefont {Q.}~\bibnamefont {Ai}},\
  }\bibfield  {title} {\enquote {\bibinfo {title} {Optical non‐reciprocity in
  coupled resonators by detailed balance},}\ }\href
  {https://doi.org/10.1002/andp.202300135} {\bibfield  {journal} {\bibinfo
  {journal} {Ann. Phys. (Berlin)}\ }\textbf {\bibinfo {volume} {535}} (\bibinfo
  {year} {2023})}\BibitemShut {NoStop}%
\bibitem [{\citenamefont {Wang}\ \emph {et~al.}(2014)\citenamefont {Wang},
  \citenamefont {Zhou}, \citenamefont {Li},\ and\ \citenamefont
  {Sun}}]{Wang2014PRA}%
  \BibitemOpen
  \bibfield  {author} {\bibinfo {author} {\bibfnamefont {Z.~H.}\ \bibnamefont
  {Wang}}, \bibinfo {author} {\bibfnamefont {L.}~\bibnamefont {Zhou}}, \bibinfo
  {author} {\bibfnamefont {Y.}~\bibnamefont {Li}}, \ and\ \bibinfo {author}
  {\bibfnamefont {C.~P.}\ \bibnamefont {Sun}},\ }\bibfield  {title} {\enquote
  {\bibinfo {title} {Controllable single-photon frequency converter via a
  one-dimensional waveguide},}\ }\href {\doibase 10.1103/PhysRevA.89.053813}
  {\bibfield  {journal} {\bibinfo  {journal} {Phys. Rev. A}\ }\textbf {\bibinfo
  {volume} {89}},\ \bibinfo {pages} {053813} (\bibinfo {year}
  {2014})}\BibitemShut {NoStop}%
\bibitem [{\citenamefont {Bliokh}\ \emph {et~al.}(2008)\citenamefont {Bliokh},
  \citenamefont {Bliokh}, \citenamefont {Freilikher}, \citenamefont
  {Savel’ev},\ and\ \citenamefont {Nori}}]{bliokh2008PRM}%
  \BibitemOpen
  \bibfield  {author} {\bibinfo {author} {\bibfnamefont {K.~Y.}\ \bibnamefont
  {Bliokh}}, \bibinfo {author} {\bibfnamefont {Y.~P.}\ \bibnamefont {Bliokh}},
  \bibinfo {author} {\bibfnamefont {V.}~\bibnamefont {Freilikher}}, \bibinfo
  {author} {\bibfnamefont {S.}~\bibnamefont {Savel’ev}}, \ and\ \bibinfo
  {author} {\bibfnamefont {F.}~\bibnamefont {Nori}},\ }\bibfield  {title}
  {\enquote {\bibinfo {title} {Colloquium: Unusual resonators: Plasmonics,
  metamaterials, and random media},}\ }\href {https://doi.org/10.1103/RevModPhys.80.1201} {\bibfield  {journal}
  {\bibinfo  {journal} {Rev. Mod. Phys.}\ }\textbf {\bibinfo {volume} {80}},\
  \bibinfo {pages} {1201} (\bibinfo {year} {2008})}\BibitemShut {NoStop}%
\bibitem [{\citenamefont {Breuer}\ and\ \citenamefont
  {Petruccione}(2002)}]{breuer2002theory}%
  \BibitemOpen
  \bibfield  {author} {\bibinfo {author} {\bibfnamefont {H.~P.}\ \bibnamefont
  {Breuer}}\ and\ \bibinfo {author} {\bibfnamefont {F.}~\bibnamefont
  {Petruccione}},\ }\href {https://api.semanticscholar.org/CorpusID:118181253}
  {\emph {\bibinfo {title} {The theory of open quantum systems}}}\ (\bibinfo
  {publisher} {Oxford University Press, USA},\ \bibinfo {year}
  {2002})\BibitemShut {NoStop}%
\bibitem [{\citenamefont {Liu}\ \emph {et~al.}(2017)\citenamefont {Liu},
  \citenamefont {Wang}, \citenamefont {Yang}, \citenamefont {Jin},\ and\
  \citenamefont {Sun}}]{Liu2017PRA}%
  \BibitemOpen
  \bibfield  {author} {\bibinfo {author} {\bibfnamefont {P.}~\bibnamefont
  {Liu}}, \bibinfo {author} {\bibfnamefont {P.}~\bibnamefont {Wang}}, \bibinfo
  {author} {\bibfnamefont {W.}~\bibnamefont {Yang}}, \bibinfo {author}
  {\bibfnamefont {G.~R.}\ \bibnamefont {Jin}}, \ and\ \bibinfo {author}
  {\bibfnamefont {C.~P.}\ \bibnamefont {Sun}},\ }\bibfield  {title} {\enquote
  {\bibinfo {title} {Fisher information of a squeezed-state interferometer with
  a finite photon-number resolution},}\ }\href {\doibase
  10.1103/PhysRevA.95.023824} {\bibfield  {journal} {\bibinfo  {journal} {Phys.
  Rev. A}\ }\textbf {\bibinfo {volume} {95}},\ \bibinfo {pages} {023824}
  (\bibinfo {year} {2017})}\BibitemShut {NoStop}%
\bibitem [{\citenamefont {Ai}\ and\ \citenamefont {Liao}(2010)}]{Ai2010CTP}%
  \BibitemOpen
  \bibfield  {author} {\bibinfo {author} {\bibfnamefont {Q.}~\bibnamefont
  {Ai}}\ and\ \bibinfo {author} {\bibfnamefont {J.~Q.}\ \bibnamefont {Liao}},\
  }\bibfield  {title} {\enquote {\bibinfo {title} {Quantum anti-{Z}eno effect
  in artificial quantum systems},}\ }\href {\doibase 10.1088/0253-6102/54/6/07}
  {\bibfield  {journal} {\bibinfo  {journal} {Commun. Theor. Phys.}\ }\textbf
  {\bibinfo {volume} {54}},\ \bibinfo {pages} {985} (\bibinfo {year}
  {2010})}\BibitemShut {NoStop}%
\end{thebibliography}

\providecommand{\noopsort}[1]{}\providecommand{\singleletter}[1]{#1}%

\end{document}